\documentclass{PoS}

\pdfoutput=1

\title{Quark Flavor Physics Review}

\ShortTitle{Quark Flavor Physics Review}

\author{\speaker{Aida X. El-Khadra}\\
        Physics Department, University of Illinois, Urbana, Illinois 61801, USA\\
        E-mail: \email{axk@illinois.edu}}

\abstract{I review the status of lattice-QCD calculations relevant to quark flavor physics. 
The recent availability of physical-mass ensembles with large physical volumes generated 
by a growing number of lattice collaborations is an exciting development and I discuss their impact on the 
landscape of lattice flavor physics calculations. The activities of the newly
formed FLAG-2 collaboration which provides averages of quantities calculated in lattice QCD
that are relevant for quark flavor physics are also discussed. My talk covers (a subset of) the same quantities reviewed 
by FLAG-2, including results for $K$, $D_{(s)}$, and $B_{(s)}$ meson decay constants and semileptonic decay 
form factors, as well as for hadronic matrix elements of neutral ($K$, $D$, and $B_{(s)}$) meson mixing. 
I also briefly discuss recent progress towards understanding nonleptonic $K$ decay and long-distance 
contributions to $\Delta m_K$.
}

\FullConference{31st International Symposium on Lattice Field Theory LATTICE 2013\\
		 July 29 -- August 3, 2013\\
		 Mainz, Germany}

\begin{document}

\section{Introduction and motivation}

Quark flavor physics has a history that is rich in important discoveries that were crucial in the 
formulation and understanding of the Standard Model (SM) of particle physics. With the recent discovery at the LHC 
of  a Higgs boson 
and with the non-observation of other new particles in the few hundred GeV to TeV energy 
range, quark flavor physics continues to be an important component of the intensity frontier and of the 
overall high energy physics program. 
The goal of the current (and planned) experimental intensity frontier effort is to constrain and possibly 
even discover beyond the SM (BSM) physics by confronting precision measurements with SM predictions. 
For many quark flavor physics processes, lattice QCD is a crucial ingredient for obtaining sufficiently precise 
SM predictions. Hence, the worldwide theoretical effort to calculate the needed hadronic matrix elements in 
lattice QCD is central to this program. 

A relevant and very recent example that illustrates the interplay between experiment and theory in 
quark-flavor physics is the rare decay $B_s \to \mu^+\mu^-$. This decay was first observed by the LHC$b$ collaboration.
Its branching fraction is now measured by LHC$b$ and CMS with a combined accuracy of 
$24\%$ \cite{CMSandLHCbCollaborations:2013pla} and we can expect the experimental accuracy to rapidly improve once
the LHC turns on again.
Recently, the Wilson coefficient for this process was calculated including three-loop perturbative QCD  
\cite{Hermann:2013kca} and two-loop EW \cite{Bobeth:2013tba} corrections, greatly reducing the 
corresponding contributions to the error budget of the SM prediction. 
Lattice QCD and CKM inputs are the dominant sources of error and 
Ref.~\cite{Bobeth:2013uxa} quotes a SM prediction with a total uncertainty of about 6\%.  
The lattice QCD input in this case is the 
$B_s$ meson decay constant (which is taken with a 2\% error) and enters into the 
SM prediction quadratically. I should also note that the CKM inputs 
depend on our knowledge of other hadronic matrix elements, calculated in 
lattice QCD, such as $B_s$ mixing parameters and $B \to D^{(*)}$ form factors. 
An alternate SM prediction can be obtained by normalizing the branching fraction to the $B_s$ mass difference, 
$\Delta m_{B_s}$ \cite{Buras:2003td}. This removes the CKM inputs and the square of the decay constant from 
the SM expression, but adds the $B_s$ mixing bag parameter. Since the bag parameter is currently known 
from lattice QCD with a precision of 
about 5\%, this alternate approach doesn't improve the precision of the SM prediction \cite{Buras:2012ru} 
at the moment. 
The $B_s \to \mu^+ \mu^-$ decay, being loop-induced in the SM, is a sensitive probe of new 
physics, and already provides strong constraints on BSM models at the current level of accuracy in both 
experiment and SM theory.

In summary,  improved lattice-QCD predictions of the hadronic matrix elements relevant for the SM predictions of  
$B_s \to \mu^+\mu^-$ can have a big impact on BSM model building, especially after the LHC turns on again, 
when the experimental measurements of the decay rate will be significantly more precise. Similar stories can be 
told for many other weak decay processes that are part of the quark flavor physics program at the intensity frontier, 
though the details are different for each process. In particular, the accuracy goals for lattice-QCD predictions of
a given quantity are dependent on the size of the experimental uncertainty as well as on the errors coming from 
short-distance or other 
phenomenologically estimated corrections (EM, isospin, EW, or perturbative QCD) which may contribute to the SM
prediction. 
The impact that a {\em broad} range of sufficiently precise lattice-QCD predictions can have on the intensity 
frontier effort cannot be overstated. As we shall see below, recent progress has us well on our way towards 
providing such predictions.

Worldwide, there are many different lattice groups (using different lattice methods) who calculate hadronic 
matrix elements relevant for a variety of weak decay processes of $K$, $D_{(s)}$, and $B_{(s)}$ mesons. With so 
many groups calculating the same matrix elements using different methods, but all providing phenomenologically 
relevant results with complete error budgets, it is desirable to combine all the lattice results for a given quantity 
into one number (and error bar). A few years ago, two such averaging efforts were started, the first by
Laiho, Lunghi, and Van de Water (LLV) \cite{Laiho:2009eu} provided averages for both heavy- and
 light-quark quantities, while the second, 
the Flavianet Lattice Averaging working group (FLAG-1) \cite{Colangelo:2010et} focused on just
light-quark quantities. Both groups computed averages  from lattice results with complete error budgets only. 
However, FLAG-1 
also provided guidance about the quality of the systematic error estimates in the form of quality criteria for the 
dominant sources of 
systematic error, including continuum extrapolation, chiral extrapolation, finite volume effects, and renormalization. 
These two efforts are now merged into one, aptly named the Flavor Lattice Averaging Group (FLAG-2), while at the same 
time increasing the membership to 28 people, in order to include researchers from as many of  the big collaborations 
working in lattice flavor physics as possible.
Like LLV, FLAG-2 now reviews heavy- and light-quark quantities, while much of the structure adopted by FLAG-1 is retained
in FLAG-2. In particular, individual working groups of three FLAG-2 members are responsible for reviewing
and averaging the lattice results for a particular quantity (or group of quantities). The quality criteria developed in FLAG-1
for the various sources of systematic error in lattice QCD calculations are also adopted in FLAG-2. 
In addition, new criteria for 
heavy-quark treatment and continuum extrapolation for the new heavy-quark quantities were developed.
In FLAG only results based on simulations with $N_f \ge 2$ flavors of dynamical fermions are considered, and separate 
averages are provided for each value of $N_f=2, 2+1, 2+1+1$. 
A preliminary version of the FLAG-2 review was posted on the FLAG website ({\tt http://itpwiki.unibe.ch/flag}) 
in Summer 2013, with the full review posted in the archives in 
October \cite{Aoki:2013ldr}. The lattice averaging effort is meant to complement the work of the Particle Data Group 
(PDG) \cite{Beringer:1900zz} and the Heavy Flavor Averaging Group (HFAG) \cite{Amhis:2012bh} who provide 
averages of experimental measurements. Combining the lattice averages together with experimental averages allows us
to obtain SM parameters and constraints on BSM theories with the best possible precision. 

The quantities that I discuss in this review are covered by four (of the seven) FLAG-2 working groups. 
In particular, my review focuses on leptonic decay constants and semileptonic form factors for kaons, $D_{(s)}$ and 
$B_{(s)}$ mesons, and on neutral ($K$, $D$, and $B_{(s)}$) meson mixing matrix elements (of local four-fermion operators). 
I also briefly mention recent results relevant for nonleptonic kaon decay as well as 
the long-distance contribution to neutral kaon mixing, neither of which are part of the current FLAG-2 effort. 
The FLAG-2 review includes lattice results made public before the end of 
April,\footnote{An update of Ref.~\cite{Aoki:2013ldr} is currently in preparation which will include results 
made public by the end of November 2013.} 
but here I also include relevant results 
presented after the end of April, at the Lattice conference, and through Fall 2013. For this purpose, the 
summary figures of Ref.~\cite{Aoki:2013ldr}, which show all the lattice calculations of a given quantity in 
comparison to the FLAG-2 averages, are modified to include these recent results for comparison purpose only. 
The FLAG-2 averages are unchanged, since these new results were not included and evaluated in 
Ref.~\cite{Aoki:2013ldr}. Nevertheless the new results give an impression of how the lattice-QCD calculations for a 
given quantity are evolving. The modified 
figures are shown in the physics sections below. 

Before discussing the physics results in Sections~\ref{sec:kaon}-\ref{sec:mix} I present a short overview of
heavy-quark methods and a discussion of the FLAG-2 criteria relevant for heavy-quark quantities in 
Section~\ref{sec:hq}. Section~\ref{sec:sim} provides a brief summary of the simulations and 
ensembles currently available for lattice quark flavor physics calculations. Both sections contain some 
introductory material which I hope will be useful to non-lattice readers. 

\section{Heavy-quark methods and FLAG-2 criteria} 
\label{sec:hq}

A detailed description of the various heavy-quark methods that are used in lattice-QCD calculations of heavy (charm 
and bottom) quark quantities is given in Appendix A.1.3 of Ref.~\cite{Aoki:2013ldr}. Here I provide a brief 
overview. 

Charm quarks can be treated with light-quark methods (provided that the action is at least $O(a)$ improved), since with 
currently available lattice spacings (see Section~\ref{sec:sim}) $am_c \gtrsim 0.15$.  A number of different light-quark 
actions are now being used for charm-quark quantities (see Section~\ref{sec:D} for more details). 
I note that an interesting new method is introduced in Ref.~\cite{Cho:2013yha}. Starting with the so-called Brillouin fermion 
action \cite{Durr:2010ch}, a tree-level $O(a^2)$ improved version
of this action is introduced in Ref.~\cite{Cho:2013yha} and tested numerically in quenched simulations for 
heavy quarks ($am_h \simeq 0.5$).\footnote{Here, and throughout this review the subscript $h$ refers
to a generic heavy quark, which may be charm or bottom, or a quark with a mass in-between the two.}
Very small cut-off effects are observed, and it will be interesting to see unquenched
calculations of charm-quark quantities with this new action, which the authors are planning to do. 

The bottom quark, however, still has $am_b > 1$ even at the smallest  lattice spacings currently available, and 
therefore cannot be treated 
with just light-quark methods.  Additional input is needed to control effects due to the large $b$-quark mass. Such input 
is provided by Effective Field Theory (EFT) in the form of  HQET  or NRQCD, and all lattice 
heavy-quark methods (designed to treat $b$-quarks)  involve the use of EFT at some stage in the calculation. The 
methods differ in how EFT is incorporated, which in turn leads to significant differences in the associated systematic 
errors between the various methods. With all methods, a numerical analysis must be based on simulations with multiple 
lattice spacings to disentangle truncation effects of the heavy-quark expansion from discretization effects and/or to 
control residual discretization effects. 

Broadly speaking, there are three classes of methods, where the first two use EFT to avoid the appearance of 
discretization errors in terms of positive powers of $(am_h)$ altogether.
In the first class of methods one starts with a continuum EFT, either HQET or NRQCD, which is then discretized 
to lattice HQET \cite{Heitger:2003nj} or lattice NRQCD \cite{Lepage:1992tx}. 
In the second class of methods, called relativistic heavy-quark actions 
(Fermilab \cite{ElKhadra:1996mp}, Tsukuba \cite{Aoki:2001ra}, 
RHQ \cite{Christ:2006us}) one uses an improved version of the Wilson action 
supplemented by guidance from HQET for the matching conditions. 
One interesting feature of relativistic heavy-quark actions is that 
they can also be used to treat charm quarks, since these
actions connect to the usual light-quark formulation in the $am_h \to 0$ limit. 
Finally, the third class of methods starts with an improved  light-quark action, 
which is used for simulations that cover a range of heavy-quark masses in the the 
charm region and up. In this case, there are of course errors which grow as 
powers of  $(am_h)$. To control them one keeps $am_h< 1$, which restricts the 
lattice data to heavy-quark masses that are smaller than the physical $b$-quark mass. 
One therefore uses HQET and possibly knowledge of the static limit to extrapolate the 
data to the physical $b$-quark mass. Two recent methods within this class are discussed in 
more detail in Section~\ref{sec:B}. 

Each heavy-quark method has its own advantages and shortcomings. For example, simulations that use lattice HQET, 
lattice NRQCD, or relativistic heavy-quark actions can be performed at the physical $b$-quark mass, without the 
need for extrapolation. But  since heavy and light quarks are treated with different actions, heavy-light currents 
must be renormalized with these methods, a step which can be a significant source of uncertainty. In addition, 
truncation errors must be considered alongside discretization errors. 
On the other hand, in simulations of heavy-quark quantities based on light-quark methods, heavy and light 
quarks are treated with the same action, and one can often use currents that are absolutely normalized. 
However, one needs to include very small lattice spacings in order to keep discretization errors under control. 

Given the variety of heavy-quark methods, a new category called "heavy-quark treatment" was 
developed for FLAG-2.  Taking into account the different issues and considerations of each heavy-quark 
method a set of minimum requirements for the level of 
improvement (and weak current matching) is specified for each approach. However, no matter what 
heavy-quark method is used, all calculations must include a careful systematic error study of truncation and/or
discretization effects.  In addition, FLAG-2 therefore adopts a data-driven quality criterion 
for ``continuum extrapolation" for evaluating lattice calculations of heavy-quark quantities. 
The purpose of the modified continuum-extrapolation criterion is to measure and evaluate
the differences between the data at the finest lattice spacings used in the simulations and the continuum 
extrapolations. Obviously, large differences would be an indication of  large residual cut-off effects. Further, 
comparing the observed deviations from the continuum extrapolated values with the quoted systematic error 
estimates provides additional 
information about the reliability of the systematic error estimates. For a detailed discussion of the  FLAG-2 
heavy-quark criteria I refer the reader to Section 2.1.2 of Ref.~\cite{Aoki:2013ldr}. 

Finally, I note that all heavy-quark methods that are 
represented in the phenomenologically relevant results discussed here (and in the FLAG-2 review) have 
been thoroughly tested, for example via calculations of heavy meson spectra and mass splittings, that can 
be compared against experimental measurements. 
A brief description of such tests is given in Section 8 of Ref.~\cite{Aoki:2013ldr}. 

\section{Status of simulations}
\label{sec:sim}

In this section I give a brief overview of the status of ensemble generation, with emphasis on simulations with
dynamical light quarks at their physical masses. There are a number of reviews
of lattice QCD (see for example Refs.~\cite{LQCDrev,Fodor:2012gf} and Appendix A of Ref.~\cite{Aoki:2013ldr})
which provide  introductions to lattice field theory and detailed descriptions of the different
choices of actions as well as of how simulations are performed. 

\begin{figure}[htb]
    \centering
    \includegraphics[width=0.6\textwidth]{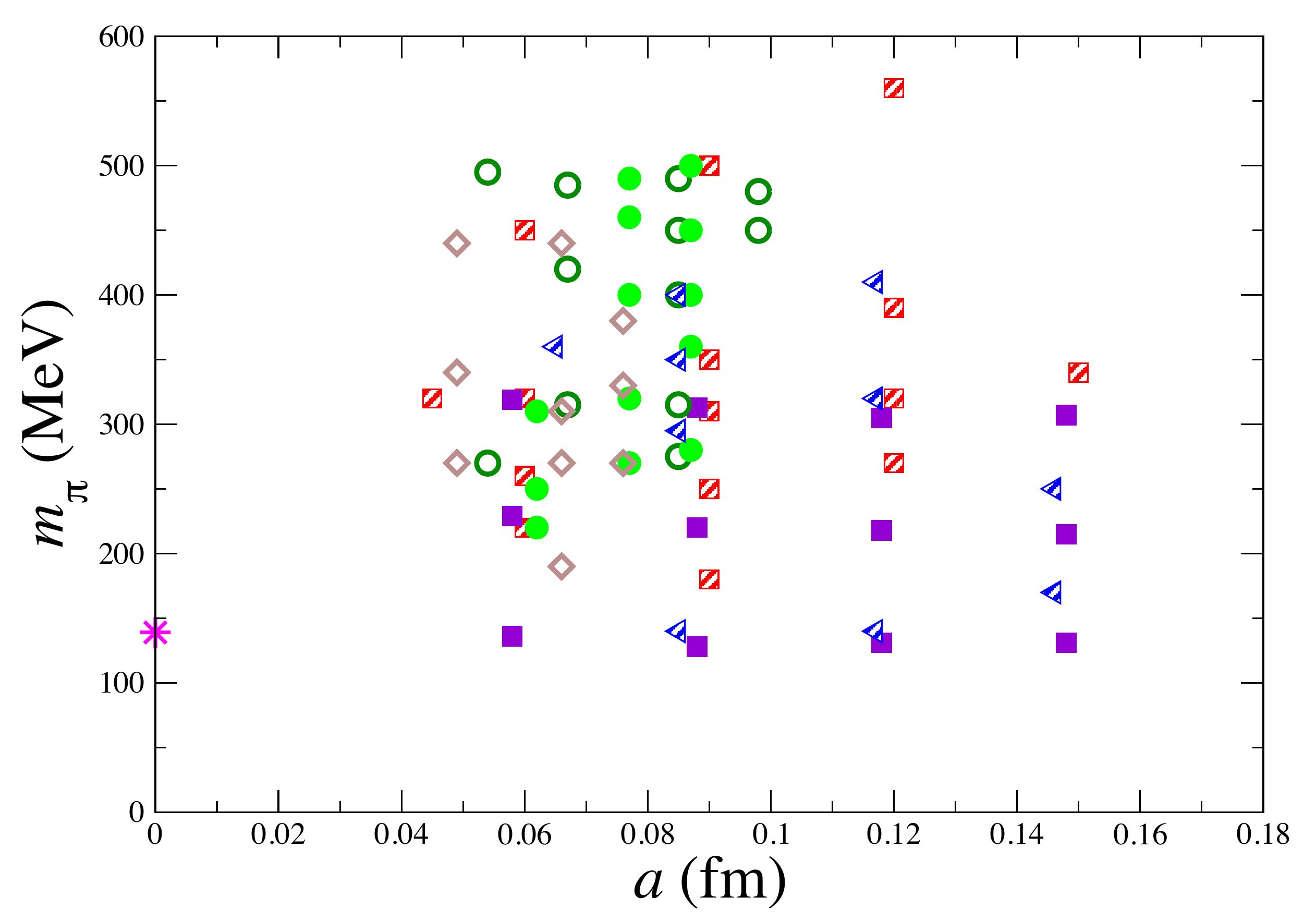}
    \caption{Ensembles used for the new results presented at Lattice 2013. The symbols mark the (Goldstone)
    sea-pion masses and lattice spacings used in a given ensemble, where simulations with $N_f = 2$, $2+1$, 
    and $2+1+1$ sea quarks are represented by open, shaded, and filled symbols, respectively.  Circles, squares,
    triangles, and diamonds indicate ensembles generated by the ETM, MILC, RBC/UKQCD, and CLS
    collaborations, respectively. The magenta burst indicates the physical point.}
    \label{fig:en}
\end{figure}

There are now close to a dozen groups who have generated ensembles with $N_f \ge 2$ flavors of 
dynamical fermions covering a range of lattice spacings, spatial volumes, and light quark masses. 
Most modern simulations use improved actions, and there is a lot of variation 
especially in the fermion actions used for the quark sea. 
In order to judge the overall quality of the simulations 
there are many aspects to consider, but the following three simulation parameters are often used as guides:
the mass of the (lightest) pion in the sea ($m_{\pi}^{\rm sea}$), the spatial extent of the lattice ($L$), 
and, of course, the lattice spacing ($a$). Since discretization effects are quantity and action dependent, the 
value of the lattice spacing itself doesn't provide a good guide to the expected discretization errors. But for 
most fermion actions with tree-level $a^2$ discretization errors, lattice-spacing errors on most light-quark 
quantities are manageable when $a < 0.1$~fm. With more improved actions larger lattice spacings can be 
used without running into large discretization errors. For heavy-quark quantities, discretization effects can
be larger than for light-quark quantities or more difficult to disentangle. A careful systematic analysis of 
discretization
effects based on simulations at multiple lattice spacings is an essential ingredient in any lattice-QCD
calculation focused on providing precision results. Indeed this has become standard practice. 
It is generally accepted that with $m_{\pi}\, L \gtrsim 4$ finite volume 
effects are under good control for light-quark quantities (involving single meson states) while 
heavy-quark quantities are generally less sensitive
to finite-volume effects. Hence, simulations with sea pions at their physical mass would benefit from having spatial 
volumes with $L \approx 6$~fm. Up until rather recently, such simulations were rare. The first
results with physical-mass sea pions were presented by the PACS-CS collaboration \cite{Aoki:2008sm} 
with $N_f = 2+1$ nonperturbatively improved Wilson flavors on a relatively small ($L\simeq 2$~fm) 
spatial volume. Since then the BMW \cite{Durr:2010vn}, MILC \cite{Bazavov:2012xda}, and 
the RBC/UKQCD \cite{Arthur:2012opa,Frison:2013fga} collaborations have generated ensembles with $N_f \ge 2+1$
with physical-mass sea pions on large spatial volumes. 
Other collaborations also have plans for (or are in the process of) generating ensembles
at the physical point. A review of the status of ensemble generation 
(as of early 2012) is provided in Ref.~\cite{Fodor:2012gf}. A nice visual summary of the 
status is shown in Figures 12 and 13 of Ref.~\cite{Fodor:2012gf}. Included in these figures are the physical-mass 
PACS-CS and BMW ensembles  but the picture is currently changing rather 
quickly, with two more groups (MILC and RBC/UKQCD) having generated ensembles at the physical point 
since early 2012, and with more to follow soon. In particular, several results obtained with the new ensembles 
generated by the MILC and RBC/UKQCD collaborations were presented at (or before) this conference. 
An overview of  all the ensembles used for the new results of quantities covered in this review that were 
presented at Lattice 2013 is given in Figure~\ref{fig:en}. 

The availability of physical-mass ensembles at small lattice spacings and on large spatial volumes represents 
an important breakthrough for lattice QCD. Chiral extrapolations become {\em interpolations}, and are used 
to correct for mis-tunings of the light-quark masses. For that purpose the inclusion of ensembles with 
larger-than-physical mass pions into an analysis based on physical-mass ensembles is still useful. In summary,
where previously the systematic error due to the chiral extrapolation and the truncation of the chiral expansion
had a prominent place in the error budget, with physical-mass simulations this error is transformed to a chiral 
interpolation error and demoted to a much less prominent place among the subleading errors. 
Not surprisingly, this development has a significant 
impact on the the overall precision that is now possible, as discussed in the physics sections below. 

A final note concerns the fact that almost all lattice groups make their ensembles available to other 
collaborations. Hence, two different results obtained using different methods may be statistically correlated 
if they use the same (or overlapping) sets of ensembles. Such correlations are taken into account in the 
FLAG-2 averages. Indeed, there are a few cases in the heavy-quark sector, where all the results to date 
have been obtained on overlapping ensembles, and taking the statistical correlations into account is 
therefore important at present.   However, once results for a given quantity are available from many 
different lattice groups, on different sets of ensembles, such correlations are less significant. This is 
indeed the case for the kaon decay constant discussed in the next section.

\section{Leptonic and semileptonic kaon decays}
\label{sec:kaon}

The focus of this section is on tree-level leptonic and semileptonic decays of kaons into 
leptonic final states of the form $\ell \nu$, which proceed via the charged-current interaction in the SM.
The study of such decays yields information on the CKM element $V_{us}$.  
While the semileptonic decay probes the $W$-boson coupling to up and down quarks via the vector 
current, the leptonic decay does so via the axial vector current. Hence observing consistency between 
the two avenues to $V_{us}$ tests the $V-A$ structure of the current. 
The hadronic matrix elements
needed to describe leptonic decays are parameterized in terms
of decay constants. Likewise, the hadronic matrix elements relevant to semileptonic decays are
parameterized in terms of form factors, see Eq.~(\ref{eq:DK}) below. 
Here I discuss the status of lattice-QCD calculations of the ratio of kaon to pion decay constants,
$f_K/f_{\pi}$, and of the semileptonic form factor $f^{K\to\pi}_+(0)$. The decay constant ratio enters the SM prediction
for the ratio of leptonic kaon to pion decay as
\begin{equation}
\frac{\Gamma ( K^+ \to \ell^+ \nu_{\ell} (\gamma) )}{ \Gamma ( \pi^+ \to \ell^+ \nu_{\ell} (\gamma) )}
= \left( {\rm known} \right)  \left| \frac{V_{us}}{V_{ud}} \right|^2 \frac{f^2_{K^+}}{f^2_{\pi^+}} 
\left( 1 + \delta^{\ell}_{\rm EM} \right)
\end{equation}
and the  form factor enters the SM prediction for semileptonic kaon decay as
\begin{equation}
\Gamma ( K \to \pi \ell \nu_{\ell} (\gamma) ) = ( {\rm known}) \left| V_{us} \right|^2 
\left| f_+^{K^0 \to \pi^-} \right|^2 (1 + \delta^{K \ell}_{\rm EM} + \delta_{SU(2)}^{K \pi} ) 
\end{equation}
where "(known)" refers to phase space and other factors that are known to high accuracy, 
and where $\delta^{\ell}_{\rm EM}, \delta^{K \ell}_{\rm EM}$ refer to structure dependent long-distance 
EM corrections which depend on the type of lepton in the final state 
(and in the semileptonic case also on the charges of the mesons).
The strong isospin breaking correction 
$\delta_{SU(2)}^{K \pi}$ is defined relative to the $K^0$ decay mode in the experimental average
of charged and neutral kaon decay. 

\begin{figure}[t]
    \centering
    \includegraphics[width=0.497\textwidth]{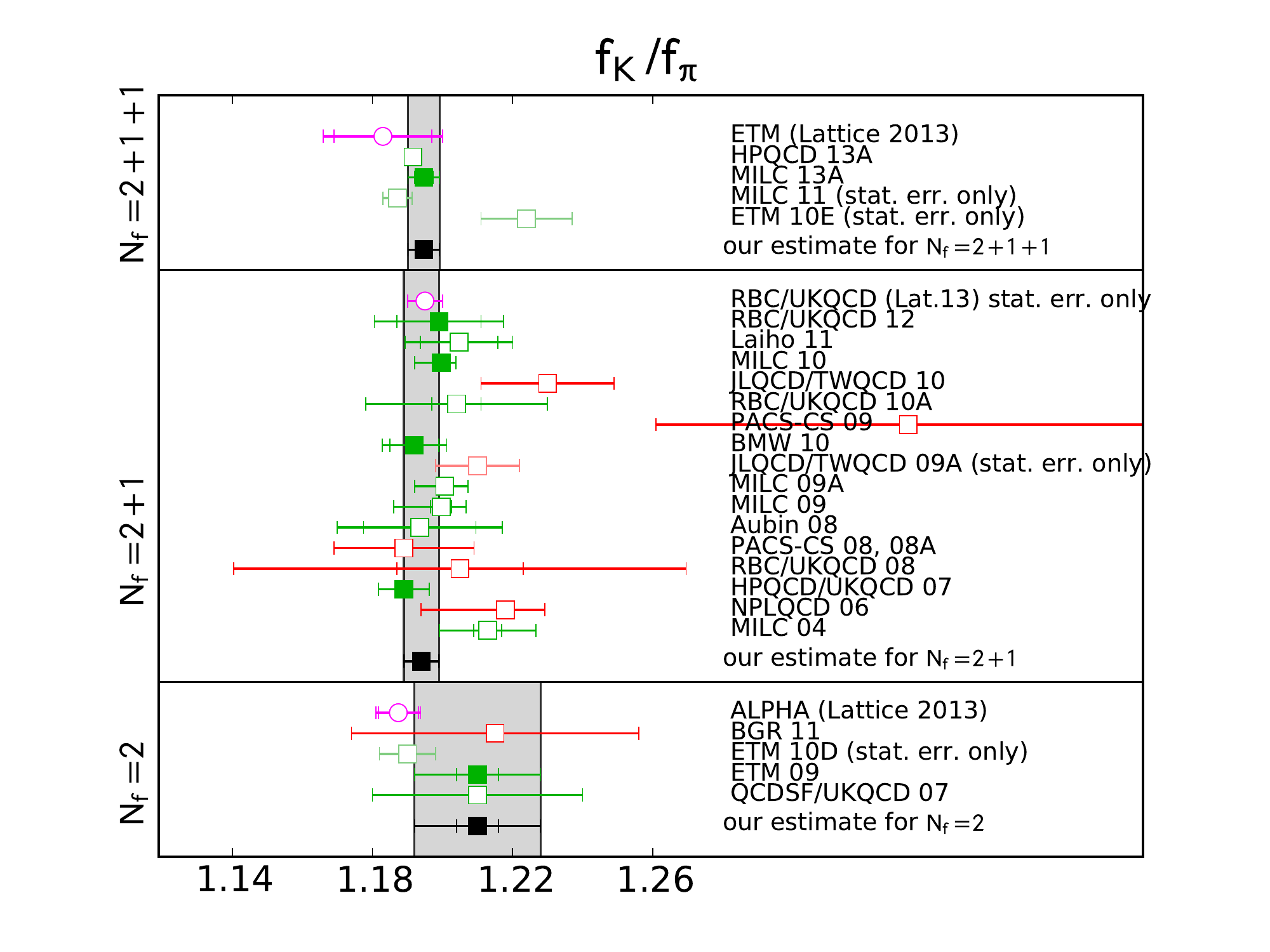} 
     \includegraphics[width=0.497\textwidth]{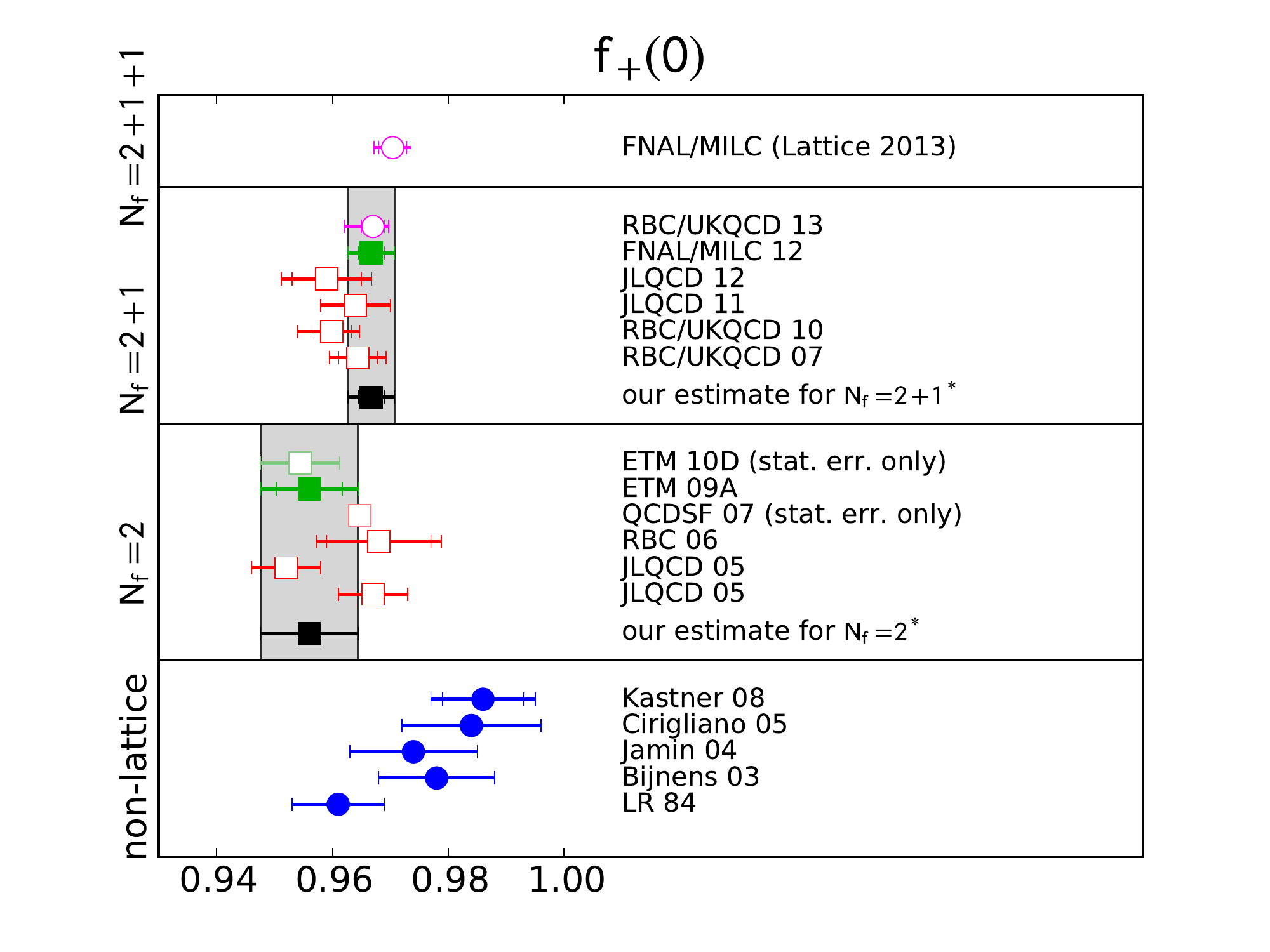}
    \caption{Overview of lattice-QCD results for (the Isospin average) $f_{K}/f_{\pi}$ (left) and for 
    $f^{K\to\pi}_+(0)$ (right) adapted from 
    Ref.~\cite{Aoki:2013ldr}. The black data points and grey bands are the FLAG-2 averages. 
    Filled green data points are results included in the respective averages, while 
    unfilled data points are not included. Blue data points (right panel) are non-lattice results.
    Magenta data points are new results presented at 
    Lattice 2013 not included in Ref.~\cite{Aoki:2013ldr} and shown here for comparison purpose only.}
    \label{fig:kaon}
\end{figure}

Typically, when a collaboration generates new ensembles, the ensemble generation often includes 
"measurements" of some simple quantities, such as pion and kaon masses (and other light hadron masses) and 
pion and kaon decay constants. The pion and kaon decay constants are therefore among the most studied flavor 
physics quantities in lattice QCD. This is illustrated in the left panel
of Figure~\ref{fig:kaon} for the ratio $f_K/f_{\pi}$ which shows a large number of entries. 
  
Naturally, lattice results exist for all values of $N_f$ for which there are ensembles, and there is no discernible difference 
(within errors) between results obtained with different $N_f$s. For $N_f = 2+1$ four lattice 
results \cite{Arthur:2012opa,Bazavov:2010hj,Durr:2010hr,Follana:2007uv} contribute to the 
FLAG-2 average, which is obtained with an uncertainty of $0.42$\%. For $N_f=2$ and $N_f=2+1+1$ there are fewer
entries, and in each case the FLAG-2 average is obtained from a single lattice 
result \cite{Blossier:2009bx,Bazavov:2013vwa}. 
In the $N_f=2+1+1$ case, the single lattice result that enters the FLAG-2 average is reported by MILC and is 
the first to be  obtained on physical-mass ensembles \cite{Bazavov:2013vwa}. 
Accordingly, a total uncertainty of $0.4$\% is quoted, smaller than for any previous result. An independent 
analysis of essentially the same data set as Ref.~\cite{Bazavov:2013vwa} 
is presented by HPQCD in Ref.~\cite{Dowdall:2013rya}, which however claims a much better precision quoting a total 
uncertainty of $0.2$\%. 
The cause of the surprisingly large error discrepancy is unclear at present. It could possibly be at least in part due to 
HPQCD using (continuum) NLO chiral perturbation theory ($\chi$PT) which may provide better control over systematics than the 
polynomial interpolations used in Ref.~\cite{Bazavov:2013vwa}. However, this is likely not the entire story, and further
study is certainly needed. In any case, even the larger error quoted by MILC in Ref.~\cite{Bazavov:2013vwa} is evidence
of the positive effect of physical-mass ensembles on the quality of lattice-QCD predictions for this quantity. 

For $N_f=2+1$ first results from physical mass ensembles have also recently appeared. A first preliminary calculation
of $f_K/f_{\pi}$ on RBC/UKQCD's physical mass ensembles was presented at this conference \cite{BMtalk}.
The analysis includes a continuum extrapolation and chiral interpolation to the physical light quark masses, but
quotes statistical errors only. The first result for $f_{\pi}$ on the $N_f=2+1$ 
physical-mass ensembles generated by the BMW collaboration is presented in 
Ref.~\cite{Durr:2013goa} with a complete error budget and a total uncertainty of 1\%, again smaller than for 
any previous result. New results for $f_K/f_{\pi}$ were also presented by the ETM collaboration \cite{Dimopoulos:2013qfa} for 
$N_f=2+1+1$ and by the ALPHA collaboration \cite{Lottini:2013rfa} for $N_f=2$.

The right panel of Figure~\ref{fig:kaon} summarizes the status of lattice-QCD results for the semileptonic form factor 
$f^{K\to\pi}_+(0)$. Here too, first results obtained on physical-mass ensembles with both 
$N_f=2+1+1$ \cite{Gamiz:2013xxa}
as well as $N_f=2+1$ \cite{Boyle:2013gsa} dynamical flavors were presented at the Lattice conference. 
FNAL/MILC \cite{Gamiz:2013xxa} quotes a total uncertainty of $0.33$\%, again smaller than any previous result. 
RBC/UKQCD \cite{Boyle:2013gsa}  quotes a larger uncertainty ($0.5$\%), because the result doesn't yet
include the new physical-mass ensembles, and the uncertainty  is therefore still dominated by chiral 
extrapolation errors.  This error will be reduced, once the physical-mass ensembles are included and
the analysis is finalized.

As discussed at the beginning of this section, both $f_K/f_{\pi}$ and $f^{K\to\pi}_+(0)$ can be used to determine 
$|V_{us}|$. At present, the lattice-QCD predictions of $f_K/f_{\pi}$ and $f^{K\to\pi}_+(0)$, 
are still the dominant sources of error in the respective $|V_{us}|$ error budgets. 
But already, Ref.~\cite{Gamiz:2013xxa} yields a 
first-row unitarity test where the contribution from the $V_{us}$ term to the total error is slightly smaller than the 
contribution from the $V_{ud}$ term. Improving the lattice-QCD calculations further is straightforward, a matter
of increasing the statistics and/or adding more ensembles. Hence, we can look forward to a time in the near future
where the dominant contributions to the $|V_{us}|$ error budgets will be from experimental uncertainties and
from the EM corrections which are currently calculated in $\chi$PT. 

A final note concerns strong isospin breaking effects. For leptonic kaon decay, the decay constant of the 
charged kaon, or, in our case $f_{K^+}/f_{\pi^+}$ is the quantity that is relevant for phenomenology. 
Some, but not all lattice groups calculate the charged meson
decay constant by extrapolating (or interpolating) to the correct valence light-quark mass. This procedure correctly
accounts for the leading isospin-breaking effects \cite{Lubicz:2013xja}. 
Other groups calculate the isospin average $f_K/f_{\pi}$, which then can be converted to $f_{K^+}/f_{\pi^+}$
either using $\chi$PT \cite{Cirigliano:2011tm} or with a separate lattice-QCD calculation based on
a new method developed in Ref.~\cite{deDivitiis:2011eh}. 
Similarly, for semileptonic kaon decay, the form factor of interest concerns $K^0 \to \pi^- \ell^+ \nu_{\ell}$ decay, which is what 
essentially all lattice groups calculate. In this case the lattice results include isospin breaking effects 
at NLO in the chiral expansion. 
The leading isospin breaking effects due to EM interactions that enter through the light-quark masses can be taken
into account by considering EM effects on light meson masses. However, the long-distance  radiative corrections 
to the leptonic or semileptonic decay ($ \delta^{\ell}_{\rm EM}, \delta^{K \ell}_{\rm EM}$) are estimated  
phenomenologically \cite{Cirig:2011xxx}. 
  
\section{Leptonic and semileptonic $D$-meson decays}
\label{sec:D}

The leptonic $D_{(s)}$-meson decay constants and semileptonic $D$-meson form factors discussed in this section
provide determinations of the CKM elements $V_{cd}$ and $V_{cs}$. 
As discussed in Section~\ref{sec:hq}, charm-quark quantities can be studied with relativistic heavy-quark
actions as well as with improved light-quark actions. Indeed, the $D_{(s)}$-meson 
decay constants have been calculated with a number of different actions. Within the improved
light-quark action category, there are results that use the HISQ action \cite{Na:2012iu,Bazavov:2012dg,Bazavov:2013nfa}, 
the twisted-mass Wilson \cite{Blossier:2009bx,Dimopoulos:2011gx,Carrasco:2013zta,Dimopoulos:2013qfa}  and 
nonperturbatively improved Wilson fermion actions \cite{Heitger:2013oaa}, and the 
Domain Wall fermion action \cite{Yang:2014taa}.
Within the relativistic heavy-quark action category, there are results that use the 
Fermilab \cite{Bazavov:2011aa} and Tsukuba \cite{Namekawa:2011wt} actions.

\begin{figure}[t]
    \centering
    \includegraphics[width=0.497\textwidth]{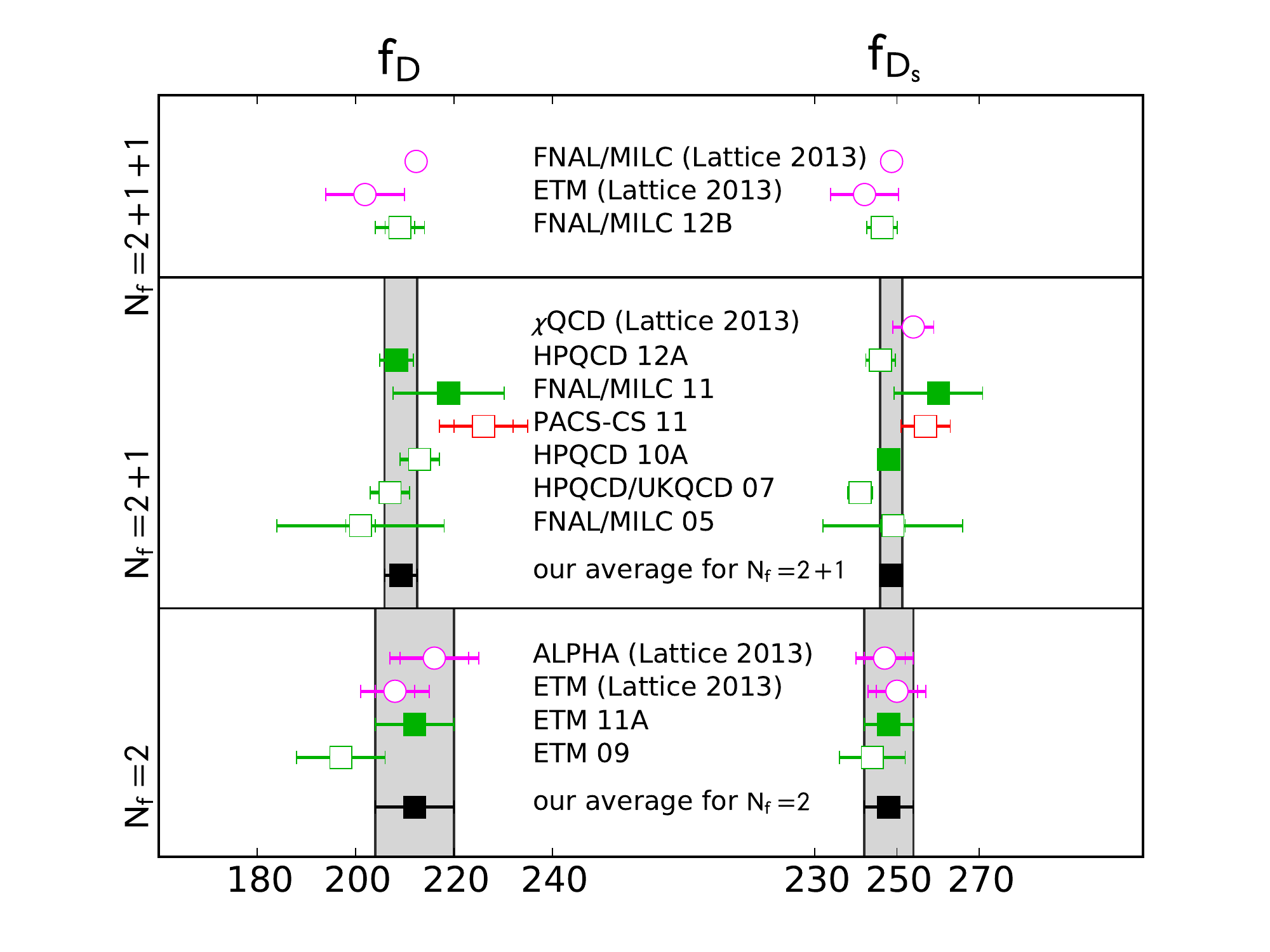} 
     \includegraphics[width=0.497\textwidth]{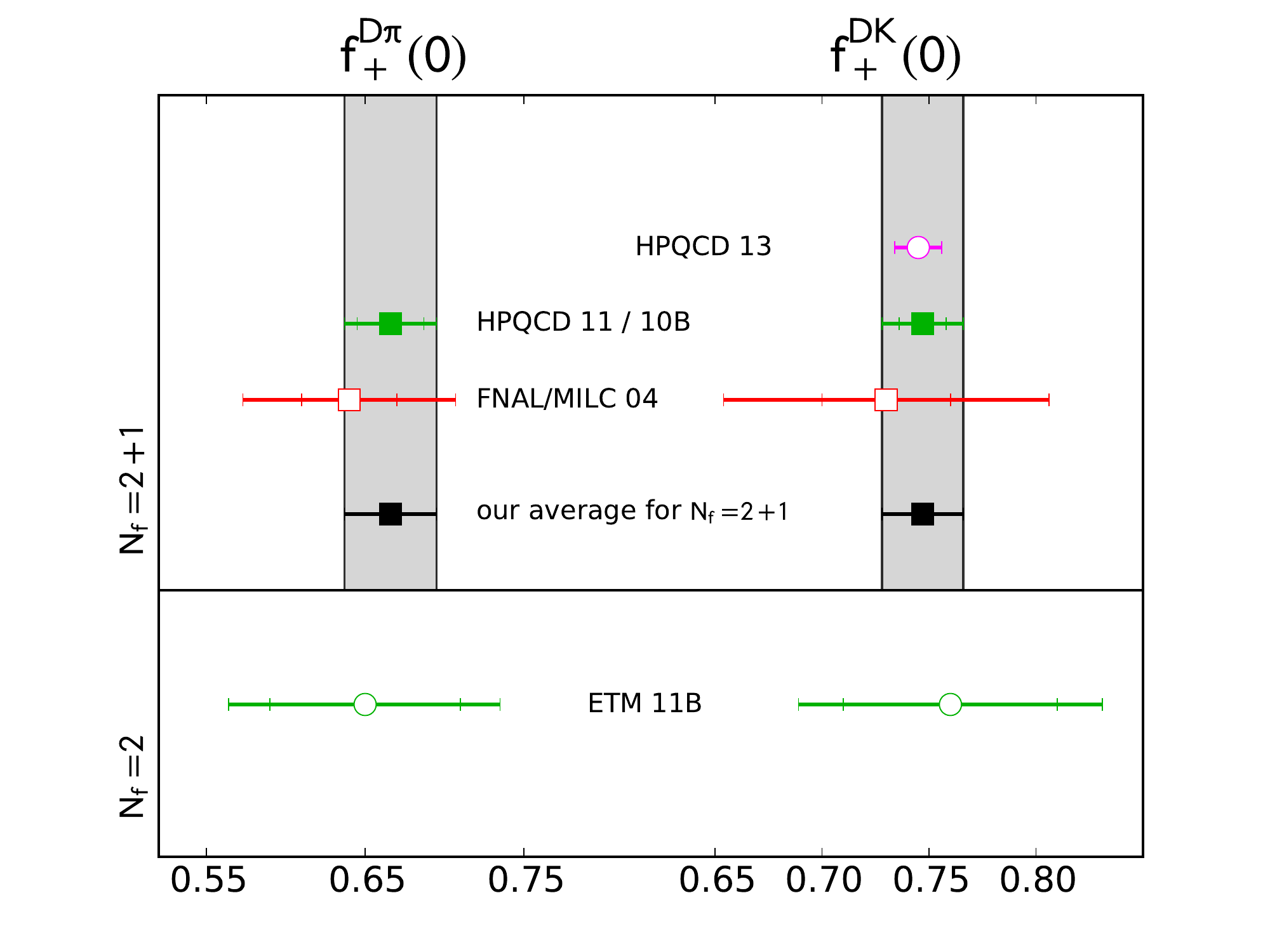}
    \caption{Overview of lattice-QCD results for $f_D$, $f_{D_s}$ (left) and $f^{D\to \pi}(0)$, $f^{D\to K}(0)$ (right) 
    adapted from Ref.~\cite{Aoki:2013ldr}. 
    See the caption of Figure~2 for an explanation of the color coding.}
    \label{fig:fD}
\end{figure}

The left panel of Figure~\ref{fig:fD} shows that despite the aforementioned variety, there are fewer independent 
results that enter the FLAG-2 averages as compared to the kaon decay constant ratio, 
two for $N_f=2+1$ \cite{Bazavov:2011aa,Na:2012iu} , one for 
$N_f=2$ \cite{Dimopoulos:2011gx}, and none for $N_f=2+1+1$ (because the result with $N_f=2+1+1$ 
\cite{Bazavov:2012dg} is a conference proceedings). For the 
$N_f=2+1$ case the FLAG-2 average is quoted with a precision of $1.1$\% ($1.6$\%) for 
$f_{D_s}$ ($f_D$), where the statistical correlations due to the two results using overlapping 
sets of ensembles are included. 
However, the left panel of Figure~\ref{fig:fD} also shows that there is a lot of recent activity, with 
five new results (shown in magenta) having been presented at the conference. 
There are two new results with $N_f=2$ dynamical 
flavors \cite{Heitger:2013oaa,Carrasco:2013zta}, one with $N_f=2+1$ \cite{Yang:2014taa} and 
two with $N_f=2+1+1$ \cite{Dimopoulos:2013qfa,Bazavov:2013nfa}. 
An update of ongoing work to extend the analysis of Ref.~\cite{Bazavov:2011aa} (which uses the Fermilab approach) 
to the complete set of  $N_f=2+1$ MILC ensembles was also reported at this conference \cite{JNStalk}. 
The analysis, which includes both the $D$- and $B$-meson decay constants, is still blind. Therefore
no quotable results were presented. The blinding factor will be removed only 
after the systematic error analysis is final. 
Finally, a preliminary calculation of $f_{D_{(s)}}$ with $N_f=2$ Domain Wall fermions was also presented 
at this conference \cite{tchiutalk}, but again without a quotable result.

The first $D_{(s)}$-meson decay constant results obtained on physical-mass ensembles were presented
by the PACS-CS collaboration \cite{Namekawa:2011wt}. However these results do not enter the FLAG-2 averages, since the 
simulations are performed at one lattice spacing only (and in a small physical volume). 
Preliminary results on physical-mass ensembles at multiple lattice spacings and on large physical volumes 
are presented by the FNAL/MILC collaborations in Refs.~\cite{Bazavov:2012dg,Bazavov:2013nfa} which use the 
MILC collaboration's $N_f=2+1+1$ HISQ ensembles.  Ref.~\cite{Bazavov:2013nfa} is an update of 
Ref.~\cite{Bazavov:2012dg} and includes ensembles at a finer lattice spacing as well as increased statistics 
on the previously used ensembles. In addition, the new analysis also includes the use of NLO Heavy Meson $\chi$PT 
(HM$\chi$PT) for all staggered fermions \cite{Bernard:2013qwa}. With all these improvements, 
the FNAL/MILC collaboration achieves a very precise result with $0.5$\% total uncertainty.  
The analysis is
published as a proceedings only, but it does include a complete and extensive error budget. 
Other factors that contribute to the smallness of the total uncertainty are the use of the
highly improved action (HISQ), which yields small discretization errors,  and of an 
absolutely normalized current, which avoids renormalization errors. Indeed, these factors also 
contributed to the precision of the results obtained by HPQCD in Ref.~\cite{Na:2012iu} which previously had the 
smallest total errors.  

A final comment concerns again strong isospin breaking effects, which are also relevant here, given
the recent results at sub-percent precision. As for kaon decay, the leading effects can simply be 
taken into account by extrapolating (or interpolating) to the correct valence light-quark mass. While most
lattice groups quote results for the isospin average $f_D$, the FNAL/MILC collaboration 
\cite{Bazavov:2011aa,Bazavov:2013nfa,Bazavov:2012dg}
calculates $f_{D^+}$. 
The experimental accuracy with which leptonic $D_{(s)}$-meson decay rates are measured suffers 
from the helicity suppression of the decay. The averages \cite{Rosner:2013ica} for the experimentally 
measured branching fractions imply contributions to the respective $|V_{cs}|$ and $|V_{cd}|$ error
budgets of $1.8$\% and $2.4$\%. 

For semileptonic $D$ (and $B$) meson decays, unlike for semileptonic kaon decay, knowledge of the form 
factors is desired over the entire kinematically allowed range of momentum transfer. 
For decays into a pseudoscalar final state, the weak hadronic matrix element can be parameterized in 
terms of the two form factors $f_+$ and $f_0$. Using the process $D \to K \ell \nu$ as an example, we have
\begin{equation}
\langle K | V^{\mu} | D \rangle = f_+(q^2) \left( p_D^{\mu} + p_K^{\mu} - \frac{m_D^2 - m_K^2}{q^2} q^{\mu} \right)
+ f_0(q^2) \frac{m_D^2 - m_K^2}{q^2} q^{\mu}\;,
\label{eq:DK}
\end{equation}
where $V^{\mu}$ is the appropriate flavor changing vector current and $q^{\mu} \equiv (p_D -p_K)^{\mu}$ is the 
momentum transfer. The SM prediction of the semileptonic 
decay rate in the rest frame of the parent meson (still using the process $D \to K \ell \nu$ as an example) reads
\begin{equation}
\frac{d \Gamma ( D \to K \ell \nu) }{d q^2} = ({\rm known}) \left| {\rm\bf p}_K \right|^3  
\left| V_{cs} \right|^2 \left| f_+^{D \to K} (q^2) \right|^2 
\end{equation}
where "(known)" is now just proportional to $G_F$. 
In order to determine, say,  $|V_{cs}|$, we just need to calculate $f^{D \to K}_+(0)$, since the 
measured differential decay rate can be extrapolated to $q^2=0$. 
However, lattice calculations of the form factor shape provide 
a test of the lattice methods via a comparison between theory and experiment. In principle, 
form factor shapes can then also be used to obtain better a determination of the corresponding CKM element. 

A summary of results for $f_+(0)$ for $D \to K$ and $D \to \pi$ semileptonic decays
is provided in the right panel of Figure~\ref{fig:fD}. Currently, only a few independent
results exist: three for $N_f=2+1$ \cite{Aubin:2004ej,Na:2011mc,Koponen:2013tua}  and 
one for $N_f=2$ \cite{DiVita:2011py}, where Ref.~\cite{Na:2011mc} forms the FLAG-2 average, 
which is quoted for only $N_f=2+1$ because the $N_f=2$ result appears in a conference proceedings. 
Refs.~\cite{Aubin:2004ej,Koponen:2013tua,DiVita:2011py} present results for  $f_{+,0}(q^2)$ (normalization and shape). 
HPQCD  calculates $f_0(q^2)$ only in Ref.~\cite{Na:2011mc}, where  the kinematic constraint $f_0(0) = f_+(0)$ is used 
to obtain the normalization, $f^{D\to K}_+(0)$ ($f^{D\to \pi}_+(0)$), with a total uncertainty of $2.5\%$ ($4.4\%$).
A new calculation by HPQCD of the $D \to K$ form factors $f_{+,0}(q^2)$ is presented in 
Ref.~\cite{Koponen:2013tua} 
for $N_f=2+1$, where a first attempt is made to determine $|V_{cs}|$ via a shape analysis that combines 
the lattice results with experimental measurements over the entire kinematic range. 
For $N_f=2$ a new calculation of the $D \to K$ and $D \to \pi$ form factors covering the entire kinematic range
was presented by the ETM collaboration  \cite{ETMDtoK} at this conference, without, however, quotable results. 
Before discussing the methods used in these two recent works in more detail, a few general comments are needed. 

As in any lattice-QCD calculation, the lattice form factors are calculated 
over a range of lattice spacings and light-quark masses, and must be 
extrapolated to the continuum and extrapolated (or interpolated) to the physical light-quark masses. But in this case,
one generates data for a set of recoil momenta in order to obtain the form factors over a certain kinematic range of recoil 
energies and there is more than one way for
performing the chiral-continuum extrapolation. A theoretically well motivated analysis strategy using a two-step 
procedure is outlined first.  
It starts with HM$\chi$PT for semileptonic decays \cite{Becirevic:2003ad,Aubin:2007mc,Bijnens:2010ws}, which, 
if supplemented
with appropriate lattice-spacing dependent discretization terms, provides a convenient framework for a combined
chiral and continuum extrapolation/interpolation. With HM$\chi$PT one parameterizes the matrix elements in terms
of the form factors ($f_{v,p}(E)$) \cite{Becirevic:2003ad} that are functions of the recoil energy and from which one 
can construct (the chiral-continuum extrapolated) $f_{+,0}(q^2)$. 

In the second step, a model-independent parameterization of the form factor shape can be obtained with so-called
$z$-expansions, which are based on analyticity and unitarity. First the momentum transfer $q^2$ is mapped to a new 
variable $z$, via
\begin{equation}
    z(q^2,t_0) = \frac{\sqrt{t_+ - q^2} - \sqrt{t+-t_0}}{\sqrt{t_+ - q^2} + \sqrt{t+-t_0}}\;,
\end{equation}
where $t_+ = (m_D+m_K)^2$ and $t_0 < t_+$ is chosen for convenience. The form factors can then written as a series
expansion in $z$,
\begin{equation}
    f(q^2) = \frac{1}{B(q^2)\Phi(q^2, t_0)} \sum_n a_n z^n \;,
    \label{eq:z}
\end{equation}
where $B(q^2)\Phi(q^2, t_0)$ contains poles and cuts, and where one can derive bounds on the sum of the squares of the
coefficients $a_n$. Different versions of $z$-expansions differ in the choice of prefactors $B(q^2)\Phi(q^2, t_0)$ 
and in how constraints are included in the coefficients. The two
most common choices are referred to as BGL \cite{Boyd:1994tt} and BCL \cite{Bourrely:2008za}.
In summary, in the two-step procedure one obtains a model-independent parameterization of the form factors 
over the desired kinematic range, where one first uses HM$\chi$PT to perform the chiral-continuum extrapolation,
and then feeds the results into $z$-expansion fits. 

In Ref.~\cite{Na:2011mc} the HPQCD collaboration introduced a modified $z$-expansion, replacing the 
two-step analysis strategy discussed above with a simpler one-step method, however, at the cost of adding ad-hoc 
assumptions about the lattice-spacing,
light-quark mass, and recoil-energy dependence of the $z$-expansion coefficients. Instead of first performing a 
chiral-continuum extrapolation, they fit their lattice form-factor data to a BCL-like function, where the $z$-expansion 
coefficients, $a_n$ are allowed to depend on the light-quark mass and on the lattice spacing
 using terms that are inspired by, but not derived from $\chi$PT and Symanzik EFT. Without
 a derivation of the modified $z$-expansion from the EFTs,  I think that detailed 
 tests are needed to understand whether or not the various modifications that are currently used  
 properly account for the systematic effects. However, a derivation of the modified $z$-expansion 
 from the EFTs would be even better. 

Ref.~\cite{Na:2011mc} uses the modified $z$-expansion  to perform a simultaneous 
chiral-continuum extrapolation of $f_0$ coupled with an interpolation to $q^2=0$. The modifications to the 
$z$-expansion coefficients include light-quark mass dependent terms, generic $a^2$ terms, heavy-quark discretization
terms, and recoil energy dependent terms. In Ref.~\cite{Koponen:2013tua} HPQCD extends the analysis of 
Ref.~\cite{Na:2011mc} to $f_{+,0}(q^2)$ over the full kinematic range. They use a subset
of the ensembles included in Ref.~\cite{Na:2011mc} (three instead of five), but increase the statistics on each 
ensemble by a factor of three or more among other improvements.  
The modified $z$-expansion used in Ref.~\cite{Koponen:2013tua} is more restricted 
with only light-quark mass and generic lattice-spacing terms added to the coefficients. The quoted result for 
$f^{D \to K}_+(0)$, with a total uncertainty of $1.5\%$ is a little less than a factor of two more precise than the 
previous result, but it is unclear what role the more restricted $z$-fit function has in this reduction of the error. 
The HPQCD collaboration also presented results for the $D_s \to \phi \ell \nu$ form factors 
\cite{Donald:2013kla}. With a vector-meson final state four form factors are needed to parameterize the matrix 
elements, and one can measure and predict three angular distributions in addition to the $q^2$ distribution. 
Ref.~\cite{Donald:2013kla} uses the same
set of ensembles and the modified $z$-expansion with the same lattice-spacing
and light-quark mass dependence as Ref.~\cite{Koponen:2013tua}. 

The ETM collaboration \cite{ETMDtoK} presents results for the $D \to K (\pi)$ form factors from two different 
analysis strategies. In the first, they perform a chiral-continuum extrapolation of the form factors at fixed $q^2$, 
where the lattice form factor 
data are first interpolated to a fixed set of $q^2$ values using a parabolic spline interpolation 
and then extrapolated to the physical light-quark
masses and to the continuum using terms linear in $m_{\pi}^2$ and in $a^2$.  In the second analysis method,
they use the modified $z$-expansion where terms constant and linear in $z$ are kept, and where the 
coefficients are extrapolated using the same terms linear in $a^2$ and $m^2_{\pi}$ as in the first method. 
The results they obtain from the two methods are consistent within the quoted errors. 
  
An almost final  comment concerns strong isospin breaking and EM effects. For semileptonic 
$D$- (and $B$-) meson decays, the experimental results are obtained from an average of charged 
and neutral meson decays, so lattice calculations of the 
isospin averages are appropriate here. However, the current HFAG averages do not take the EM correction 
(of order $2\%$) due to the Coulomb attraction between the charged final states in neutral meson semileptonic 
decays into account.  
Finally, I comment on how form factor results are presented, and what is needed when a calculation quotes more
than just one number (and error budget). 
If a work presents results for both normalization and shape, it also should provide the corresponding covariance matrix, 
in addition to a complete error budget, so that the results can be properly combined with other lattice calculations and 
with experimental results.

\section{Leptonic and semileptonic $B_{(s)}$-meson decays}
\label{sec:B}

The leptonic $B$-meson decay constants and semileptonic $B$-meson form factors discussed in this section
provide determinations of the CKM elements $V_{ub}$ and $V_{cb}$. However, the quantities discussed in this section
can also be used to obtain SM predictions of rare processes, such as the $B_s \to \mu^+ \mu^-$ decay already 
discussed in the introduction, that are sensitive to contributions from new physics and can therefore be used to 
constrain BSM theories. 

Before presenting the lattice results, I would like to discuss two relatively new heavy-quark methods 
that use light-quark actions (with $O(am_h)^2$ errors) in calculations of  $b$-quark quantities. 
The ETM collaboration has developed an interesting strategy, aptly named the ``ratio method'' \cite{Blossier:2009hg}, 
in which they form ratios of physical quantities such that the static limit of the ratios is equal to unity and where discretization 
errors are suppressed by the step-scaling factor used in constructing the ratios. Using decay constants as an example, 
and denoting the decay constant for a meson with heavy quark flavor $h$ and light valence quark flavor $\ell$ as $f_{h\ell}$, 
and the corresponding heavy-meson mass as $M_H$, it is well known that the combination 
$\phi(m_h) \equiv f_{h\ell} \sqrt{M_H}$ 
has a well-defined static limit ($m_h \to \infty$).\footnote{In their main analysis, ETM actually 
uses the heavy-quark pole mass instead of the 
heavy-meson mass for $\phi$.} 
Then the ratio 
\begin{equation}
z_{h\ell}(m_h) \equiv \frac{\phi(m_h)}{\phi(m_h/\lambda)} \;
\label{eq:zdef}
\end{equation}
(with $\lambda$ a fixed step-scaling parameter) has the property that $z(m_h \to \infty) \to 1$. The $z$-ratios 
have the further property that heavy-quark discretization 
errors are suppressed with a suitable choice of step-scaling factor $\lambda$.  ETM then uses the twisted-mass 
Wilson action to calculate the  $z$-ratios
 for a series of fixed heavy-quark masses 
(separated by the step scaling factor $1 < \lambda \lesssim 1.3$) starting from the charm quark mass up to some 
heavy-quark mass while keeping $am_h < 1$. In addition they calculate the physical quantity of interest 
($\phi$ in this example) 
at the charm-quark mass. After chiral-continuum extrapolation of the $z$-ratios 
(and $\phi(m_c)$) they then interpolate the $z$-ratios to the physical $b$-quark mass using guidance from HQET, 
and finally construct the quantity of interest ($f_B$ in this case) from the step-scaled product of $z$-ratios and 
$\phi(m_c)$.  This procedure, outlined here for the case of heavy-meson decay constants can be applied to any
heavy-quark quantity that has a well-defined static limit. Indeed, ETM has used this method in determinations 
of the $b$-quark mass, as well as in 
calculations of $B_{(s)}$-meson decay constants  and of $B_{(s)}$-meson mixing 
parameters \cite{Carrasco:2013zta,Carrasco:2013naa} (see Section~\ref{sec:mix}). In addition the Orsay 
group has used this method for calculations of semileptonic form factors for 
$B_{(s)} \to D^{(*)}_{(s)}$ decays \cite{Atoui:2013sca,Atoui:2013ksa,Atoui:2013zza} . 
Finally, I note that ETM's ratio method can be used with any light-quark action that has $O(a^2)$ discretization 
errors. The cancellation of heavy-quark discretization errors observed in the $z$-ratios is a generic feature of 
the ratios themselves, which I would expect to be largely independent of the underlying light-quark action. 

Another interesting method, presented by the HPQCD collaboration in Ref.~\cite{McNeile:2011ng}, is  
dubbed the ``heavy HISQ'' method. This method takes advantage of the fact that the leading heavy-quark 
discretization errors are suppressed in the HISQ action \cite{Follana:2006rc} compared to other 
light-quark actions, a fact which has been tested extensively in numerical simulations \cite{McNeile:2012qf}. 
The HISQ action, like the Asqtad 
action, is essentially a tadpole-improved staggered action (albeit with much reduced taste-violating effects), 
which means that the leading heavy-quark discretization errors are $O(\alpha_s (am_h)^2, (am_h)^4)$. 
However, in the construction of the HISQ action, the leading heavy-quark discretization errors are further
reduced by factors of $v/c$ via an adjustment of the Naik term, yielding leading errors of the form:
\begin{equation}
 \alpha_s \, (v/c) (am_h)^2 = \alpha_s \, (a \Lambda) (a m_h)  \;\;\;\;\;\;\;\;\;
 (v/c)^2 (am_h)^4 =  (a \Lambda)^2  (a m_h)^2 \;,
\end{equation}
where the right sides of the two equations apply to heavy-light systems with $\Lambda$ as the 
typical momentum of the heavy quarks inside the heavy-light mesons. Another 
important ingredient in this approach, both for studying and controlling discretization effects and for 
reaching large enough heavy-quark masses (in physical units), is the availability of ensembles 
that cover a large range of lattice spacings and reach down to very small lattice spacings. The 
HPQCD collaboration used the MILC collaboration's
Asqtad ensembles (shown as shaded red squares in Figure~\ref{fig:en}), which include lattice spacings in the
range $a \approx 0.045 - 0.15$~fm.  In Ref.~\cite{McNeile:2011ng} the $f_{B_s}$ decay
constant is calculated from relativistic data that include heavy-quark masses as large as $am_h \lesssim 0.85$. 
Surprisingly they find that discretization errors (as measured by the observed deviations from the 
continuum fit line) decrease with increasing heavy-meson mass $M_H$ (in physical units). They perform a 
combined continuum extrapolation and HQET-inspired heavy-quark extrapolation to the physical $b$-quark
mass. The heavy HISQ method has also been used in studies of heavy-light and heavy-heavy meson spectra
and decay constants \cite{McNeile:2012qf} as well as current-current correlators \cite{McNeile:2010ji}.
Chiral extrapolations are avoided in these calculations by focusing on heavy-quark quantities with 
strange (or heavier) valence quarks where sea-quark mass dependence is a subdominant source of error 
(at the current level of precision).
 
 \begin{figure}[t]
    \centering
    \includegraphics[width=0.497\textwidth]{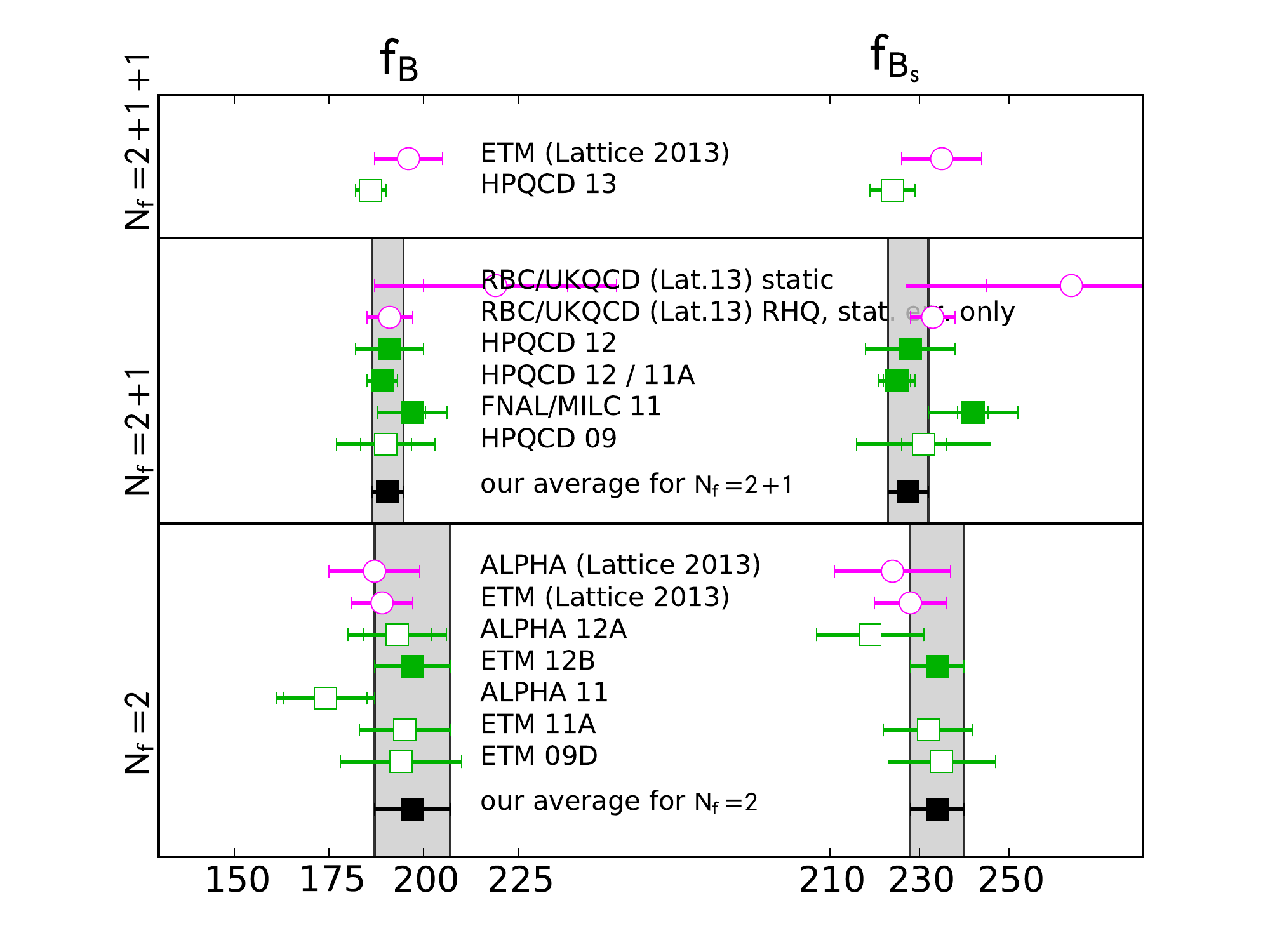}
        \includegraphics[width=0.497\textwidth]{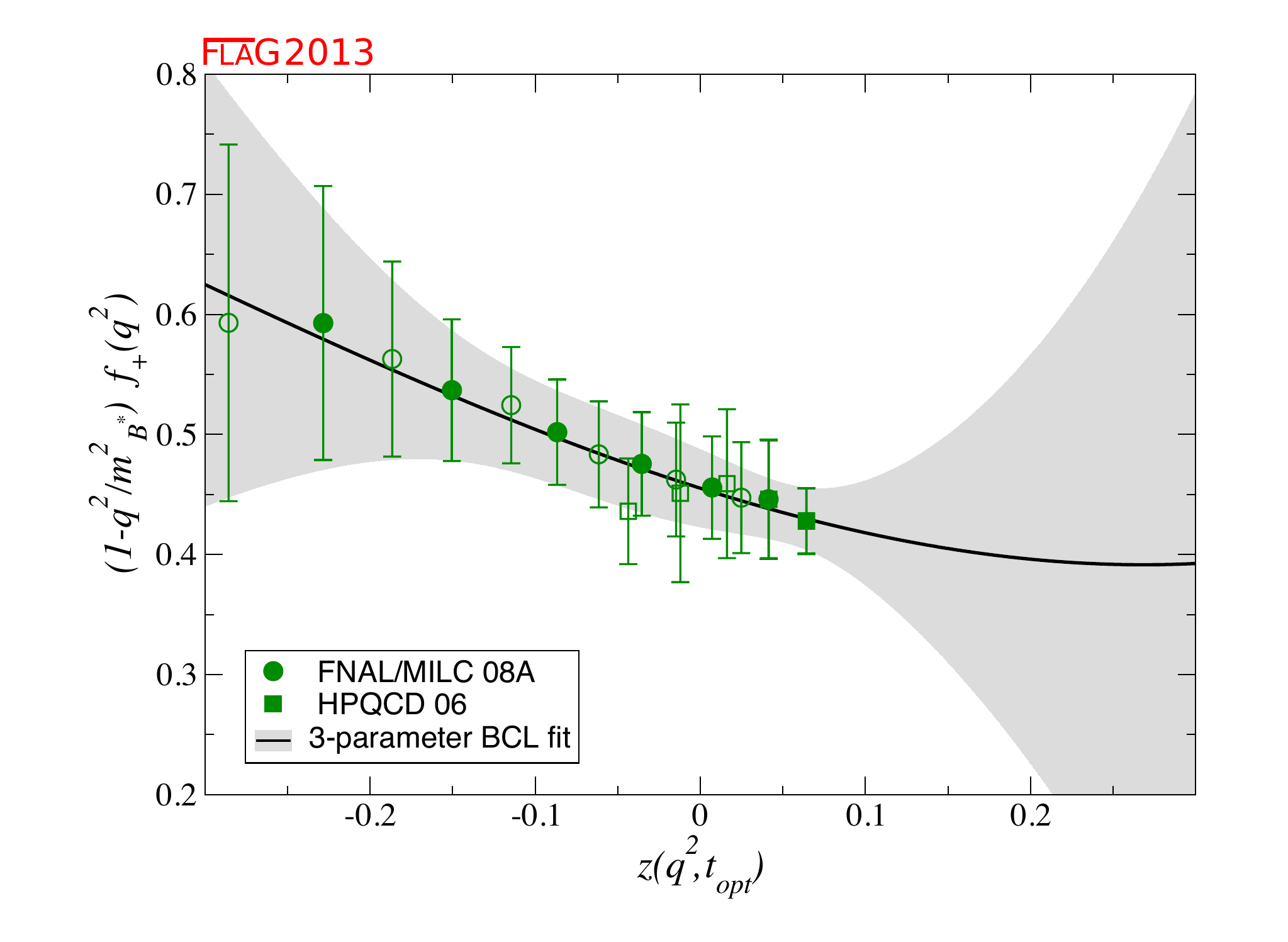} 
    \caption{Overview of lattice-QCD results for $f_B$, $f_{B_s}$ (left) and $(1-q^2/m_{B^*}^2) f^{B \to pi}_+(q^2)$
  (right). The plot in the left panel is adapted from Ref.~\cite{Aoki:2013ldr} while the right panel is taken directly
  from Figure 20 of Ref.~\cite{Aoki:2013ldr}. See the caption of Figure~2 for an explanation of the color coding 
  in the left panel. For the right panel, filled symbols are lattice data included in the 3-parameter BCL fit, open 
  symbols are not. The FLAG-2 fit curve is shown as a black line with grey error band.}
    \label{fig:fB}
\end{figure}
The left panel of Figure~\ref{fig:fB} shows that the status of $B$-meson decay constant calculations is 
similar to the case of $D$ mesons, both for considering the number of lattice results that enter the FLAG-2 
averages for $f_{B_{(s)}}$, and for considering the number of new results presented at this conference. 
Another similarity is that the $B$-meson decay constant results are obtained with a variety of different heavy-quark
methods, including the ratio method (with twisted-mass Wilson fermions) 
\cite{Carrasco:2012de,Carrasco:2013zta,Carrasco:2013naa}, the heavy-HISQ method \cite{McNeile:2011ng},
lattice HQET \cite{Bernardoni:2013oda}, lattice NRQCD \cite{Na:2012kp,Dowdall:2013tga}, 
Fermilab heavy quarks \cite{Bazavov:2011aa}, the RHQ action \cite{Witzel:2013sla}, 
and the static limit \cite{Ishikawa:2013faa}. 
However, the uncertainties on both the individual lattice results and on the FLAG-2 averages are larger in the case
of $B$-meson quantities, reflecting the larger discretization/truncation errors and also the generally larger
statistical errors. 
Three results enter the FLAG-2 averages for $N_f=2+1$ \cite{Bazavov:2011aa,McNeile:2011ng,Na:2012kp} , one for 
$N_f=2$ \cite{Carrasco:2012de}, and none for $N_f=2+1+1$ 
(because the $N_f=2+1+1$ result of Ref.~\cite{Dowdall:2013tga} was not published in a journal until after the
FLAG-2 April 30 deadline). For the 
$N_f=2+1$ case the FLAG-2 average is quoted with a precision of $2.0$\% ($2.2$\%) for 
$f_{B_s}$ ($f_B$), where the statistical correlations due to the three results using overlapping 
sets of ensembles are included.
Five new results (shown in magenta in the left panel of Figure~\ref{fig:fB}) were presented at this conference, 
two with $N_f=2$  \cite{Carrasco:2013zta,Bernardoni:2013oda}, two with 
$N_f=2+1$ \cite{Witzel:2013sla,Ishikawa:2013faa}  and one with 
$N_f=2+1+1$ \cite{Carrasco:2013naa}. 

The first $B_{(s)}$-meson decay constant results obtained on physical-mass ensembles are presented by the
HPQCD collaboration in Ref.~\cite{Dowdall:2013tga} on a subset of the MILC collaboration's $N_f=2+1+1$ HISQ 
ensembles.  However, here, the gain from the use of physical-mass ensembles is less apparent, since
for $B$-meson quantities there typically are other sources of uncertainty that dominate (or significantly 
contribute to) the error budget.  In this 
particular case, the dominant error is
due to the renormalization of the NRQCD-HISQ current, calculated at 
one-loop order in mean-field improved perturbation theory. A comparison of the error budgets between this result 
and HPQCD's previous calculation with NRQCD-HISQ valence quarks 
\cite{Na:2012kp} performed on a set of Asqtad ensembles over roughly the same range of lattice spacings, 
shows a reduction in the total uncertainty by roughly a factor of two.  However, a closer look at the error budgets
reveals that the biggest change is in the perturbative error estimates. Since the valence-quark actions and the 
lattice spacings are the same in both cases, so too are the perturbative renormalization factors the same
(up to small differences due to the masses being slightly different in the two cases). However, 
Ref.~\cite{Dowdall:2013tga} quotes a perturbative uncertainty that is smaller by almost a factor of three. They
justify this error reduction by reorganizing the perturbative expansion in a manner consistent with the 
mostly NPR method \cite{Harada:2001fi}. 

Because of the helicity suppression the only experimentally measured
leptonic $B$-meson decay rate is for the $B \to \tau \nu_{\tau}$ channel. The branching ratio measurements
still have rather large errors and the experimental average of Ref.~\cite{Rosner:2013ica}  implies a 
contribution to the $|V_{ub}|$ error budget of around $10$\%. 
  
For semileptonic decays of $B_{(s)}$ mesons, we distinguish between light final states 
(heavy-to-light case) and charm final states (heavy-to-heavy case), which we will discuss in turn.
For the heavy-to-light case, we further distinguish between tree-level weak decays (such as 
$B \to \pi \ell \nu$ or $B_s \to K \ell \nu$) that are relevant to
$|V_{ub}|$ determinations and rare (or FCNC) decays (such as $B \to K^{(*)} \ell^+ \ell^-$) that are 
used to constrain BSM theories. The SM decay rates for semileptonic $B$-meson decays to 
$\tau$-leptons receive a contribution from the scalar form factor $f_0(q^2)$ as well as the usual
 $f_+(q^2)$, while rare semileptonic $B$-meson decay rates also receive contributions from the 
 tensor form factor $f_T(q^2)$. 
For semileptonic $B$-meson decays to light hadrons the recoil momentum range accessible to 
lattice-QCD calculations is significantly smaller than the full kinematic range of the decay. 
Furthermore, experimental measurements of differential decay rates are most accurate in the 
high recoil momentum region not directly accessible to lattice QCD. Model-independent 
parameterizations of the form factor shape are essential for obtaining information about the form 
factors outside the lattice-data region and all modern lattice-QCD calculations use 
$z$-parameterizations for just that purpose. 

The FLAG-2 review includes averages only for $N_f=2+1$ and only for the $B \to \pi$ channel at present, 
since this was the only channel with published lattice results at the time the review was written. The right panel of
Figure~\ref{fig:fB} illustrates the status of $B \to \pi$ form factor calculations prior to this conference. The two 
calculations \cite{Dalgic:2006dt,Bailey:2008wp}  that are included in the FLAG-2 form factor average are both  
over five years old. The first \cite{Dalgic:2006dt} uses lattice NRQCD to treat the $b$ quark while the second
\cite{Bailey:2008wp} is based on the Fermilab approach. Only Ref.~\cite{Bailey:2008wp} quotes the covariance 
matrix for the form factor results, and the FLAG-2 average is therefore obtained from combining the data from  
Ref.~\cite{Bailey:2008wp} with one data point from Ref.~\cite{Dalgic:2006dt}. Since the two results use 
overlapping sets of ensembles statistical correlations are included in the average. 
FLAG-2 performs a (constrained) BCL fit to the combined lattice data, which is shown as a grey band 
in the right panel of Figure~\ref{fig:fB} and then further combines the lattice data with experimental data
from BaBar \cite{Lees:2012vv} and Belle \cite{Ha:2010rf}, without, however, combining both experiments. 
This procedure yields an uncertainty on $|V_{ub}|$ in either case of about $6$\%, where the 
longstanding tension between the $|V_{ub}|$ determinations from exclusive and inclusive  semileptonic
$B$ decays still stands at about $3 \sigma$. 
The good news is that a number of new calculations for this channel are underway both for the $N_f=2+1$ 
\cite{Du:2013kea,Bouchard:2013zda,Kawanai:2013qxa} and $N_f=2$ \cite{Bahr:2012vt} cases. Further good news
is that each of these projects is based on a different heavy-quark action, including Fermilab, 
NRQCD, RHQ, and HQET. Some groups \cite{Bouchard:2013zda,Liu:2013sya}
are also now calculating the form factors for $B_s \to K \ell \nu$ for $N_f=2+1$.
This process has not yet been measured experimentally. Once it is, it will provide an alternate determination
of $|V_{ub}|$. All the new results (or calculations in progress) described here use the theoretically well-motivated
two-step analysis strategy outlined in the previous section, which allows for complete control over the dominant
systematic errors over the full kinematic range, provided, of course, that the underlying numerical simulations
cover a large enough range of lattice spacings and light-quark masses. 

Rare semileptonic decays have also received attention recently. The first published $N_f=2+1$
results for the rare decay $B \to K \ell^+ \ell^-$ form factors appeared by the HPQCD collaboration
this summer \cite{Bouchard:2013eph},
where the form factors are calculated with $\approx 5-9\%$ uncertainty in the low-recoil region. 
Another calculation of this process by the FNAL/MILC collaboration
is also close to completion \cite{Zhou:2012dm,Liu:2013sya}. 
Ref.~\cite{Bouchard:2013eph} compares the $z$-fit results for the form factors from the two-step 
analysis to results obtained from a modified $z$-expansion fit (see Section~\ref{sec:D}).  Overall consistent 
behavior between the two fit bands is seen, even in the extrapolated region, but the coefficients of 
the two $z$-fits are not compared in detail.  
Ref.~\cite{Bouchard:2013eph} (second paper) also provides the SM prediction for the differential
decay rate implied by HPQCD's form factor results which is in turn compared to experimental measurements
showing overall good agreement. 

Rare semileptonic $B$-meson form factors for decays to 
vector final states, i.e. $B \to K^* \ell^+ \ell^-$ and $B_s \to \phi \ell^+ \ell^-$, are calculated by the Cambridge
group in Ref.~\cite{Horgan:2013hoa}. With vector final states there are a number of interesting observables 
that can be studied in detailed comparisons between theory 
and experiment. The second paper in Ref.~\cite{Horgan:2013hoa} provides SM predictions as well 
as constraints on the BSM contributions to the Wilson coefficients $C_9$ and $C'_9$. 
As is well known, the lattice-QCD methods currently used for studying 
weak decays do not properly treat final states that are resonances, such as the $K^*$ and $\phi$ 
mesons. An exception is the $D^*$ meson (see below),
where $\chi$PT guides the chiral extrapolation to the physical light-quark masses. 
The calculation of Ref.~\cite{Horgan:2013hoa}, while providing a study of other systematic
errors, doesn't include an estimate of this source of error. One can reasonably argue that the 
small width of the $\phi$ meson may imply a smaller systematic effect in the $B_s \to \phi$ channel, 
but without further study to quantify the effect one has to keep in mind that the SM predictions 
and BSM constraints derived from such lattice form factor
calculations contain an unknown systematic error. 
This situation is, however, no worse than for  
phenomenological approaches to form factor calculations based, for example, on light-cone 
sum rules \cite{Ball:2004rg}.

 \begin{figure}[t]
    \centering
    \includegraphics[width=0.55\textwidth]{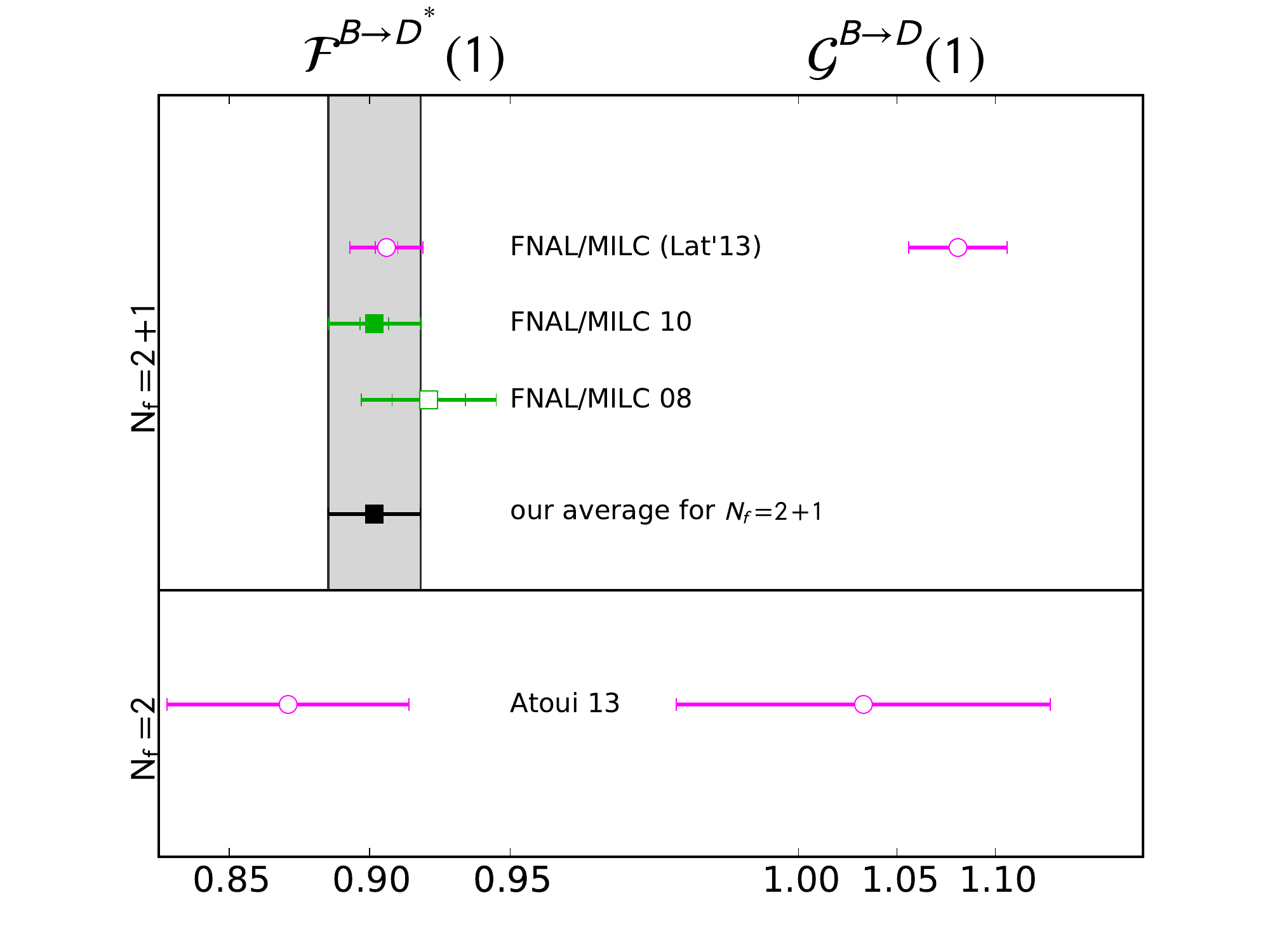} 
    \caption{Overview of lattice-QCD results for ${\cal F}^{B\to D^*}(1)$, ${\cal G}^{B\to D}(1)$ 
    adapted from Ref.~\cite{Aoki:2013ldr}. See the caption of Figure~2 for an explanation of the color coding.}
    \label{fig:BtoD}
\end{figure}
The current FLAG-2 average for the $B \to D^*$ form factor at zero recoil  (available only for $N_f=2+1$ at present)
is based on a series of calculations by the FNAL/MILC collaboration \cite{Bernard:2008dn} with Fermilab bottom 
and charm valence quarks. The results
are summarized in 
Figure~\ref{fig:BtoD}. The observed reduction in error from the initial result in 
Ref.~\cite{Bernard:2008dn} quoted with a $2.6$\% uncertainty to the most recent result presented in 
Ref.~\cite{Qiu:2013ofa} with a quoted uncertainty of $1.4$\% reflects improvements due to adding ensembles
at smaller lattice spacings, with smaller light-quark masses, and with increased statistics. The final result of this series
of calculations presented in Ref.~\cite{Qiu:2013ofa} is based on essentially the full set of Asqtad ensembles
with lattice spacings in the range of $a \approx 0.045 - 0.15$~fm. The finite (but small) width of the $D^*$-meson
can be treated within the framework of HM$\chi$PT, and yields a cusp in the continuum-extrapolated chiral-fit curve. 

The lattice-QCD result for the $B \to D^*$ form factor reported in Ref.~\cite{Qiu:2013ofa} can be used to determine
 $|V_{cb}|$ from the experimental average \cite{Amhis:2012bh} with a total uncertainty of $1.9$\%. For the first time, 
 the contribution from the lattice form factor uncertainty to the $|V_{cb}|$ error budget is commensurate with the 
 contribution from the experimental uncertainty. 
The experimental average in Ref.~\cite{Amhis:2012bh} does not correct for the Coulomb attraction 
between the charged final state particles 
in the neutral $B$-meson decays. At this level of precision such corrections need to be included, and this is 
done in the $|V_{cb}|$ determination reported in Ref.~\cite{Qiu:2013ofa}. 
The longstanding tension between the determination
of $|V_{cb}|$ from exclusive and inclusive decays currently stands at about $3 \sigma$. Given this tension, and given
the overall importance of $V_{cb}$ in normalizing the unitarity triangle and as a parametric source of uncertainty
in the SM model predictions of processes such as $B_s \to \mu^+\mu^-$, 
$K \to \pi \nu \bar{\nu}$,  and neutral kaon mixing ($\epsilon_K$), it is important to have additional lattice-QCD
calculations, both of the same form factor and of the related $B \to D$ form factor, preferably using different 
heavy-quark methods. Fortunately there are
four new efforts at various stages of completion, three of them independent of Ref.~\cite{Bernard:2008dn}. 
However, more would be welcome. 

The Orsay group \cite{Atoui:2013sca} reports a very preliminary result for the $B \to D^*$ form factor 
shown in 
Figure~\ref{fig:BtoD}. They use the ratio method and the twisted-mass Wilson action
to calculate the form factor  at two lattice spacings on $N_f=2$ ensembles generated by the ETM collaboration. 
The main focus of this study  is on the form factors for $B \to D^{**}$ decays (i.e. to orbitally 
excited $D$-mesons) \cite{Atoui:2013ksa}, which may play a role in resolving the tension between 
exclusive and inclusive $|V_{cb}|$ determinations. The reported  $B \to D^*$ form factor result provides a 
consistency check of the method used to determine the $B \to D^{**}$ form factors. 
The Orsay group \cite{Atoui:2013zza} also presents a calculation of $B_{(s)} \to D_{(s)}$ form factors 
using again the ratio method with twisted-mass Wilson fermions and using  $N_f=2$ ensembles 
generated by the ETM collaboration at four different lattice spacings in the range 
$a \approx 0.054 - 0.098$~fm. They calculate ${\cal G}^{B_s \to D_s}(1)$ (${\cal G}^{B \to D}(1)$) with a
precision of about $4$\% ($9.2$\%) 
and also determine the scalar and tensor form factors near zero recoil . The scalar form factor contributes
to semileptonic decays to $\tau$-lepton final states and both form factors may provide constraints on BSM models. 
The Orsay group finds results that are consisted with those reported in Ref.~\cite{Bailey:2012xxx}. 

Ref.~\cite{Qiu:2013ofa} presents a $N_f=2+1$ calculation of the $B \to D$ form factors at non-zero 
recoil. It uses the same valence-quark actions and the same ensembles as the $B \to D^*$ form factor analysis. 
The $B \to D$ form factors are calculated for the full kinematic range using a two-step analysis (as described 
in Section~\ref{sec:D}) with a BGL $z$-expansion fit.  A complete error budget is provided, and $|V_{cb}|$ is 
determined from a combined $z$-fit of the lattice form factors together with experimental data by 
BaBar \cite{Aubert:2009ac} with a total uncertainty of about $5$\%. The analysis was kept blind until the
systematic error budget was final, and the result for $|V_{cb}|$ is nicely consistent with the one from 
the $B \to D^*$ analysis, albeit less precise.  
The HPQCD collaboration is in the process of calculating
the $B_{(s)} \to D_{(s)}$ form factors using NRQCD $b$ quarks and HISQ charm 
quarks \cite{Bouchard:2013zda}. In another new effort \cite{Jang:2013yqa}, it is planned to use the 
highly improved OK action 
\cite{Oktay:2008ex} for a calculation of the $B \to D^{(*)}$ form factors since heavy-quark discretization 
errors are a dominant source of uncertainty with the Fermilab method. This project has just been started. 

\section{Neutral ($K$, $D$, and $B_{(s)}$) meson mixing and nonleptonic kaon decay}
\label{sec:mix}

Neutral meson mixing, being loop-induced in the SM, plays an important role in determining the 
CP violating parameters of the SM as  well as in providing constraints on BSM theories. 
In the SM, neutral meson mixing receives contributions from hadronic matrix elements of 
$\Delta F = 2$ local operators, (where $F=B,C,S$). 
For neutral kaons and  $B_{(s)}$-mesons these are the dominant contributions to the SM predictions of 
the corresponding mixing observables ($\epsilon_K$, $\Delta m_{d(s)}$). In BSM theories, additional
$\Delta F = 2$ local operators can contribute, and the most general $\Delta F = 2$ effective hamiltonian can be written
in terms of five  operators,
\begin{equation}
{\cal H}_{\rm eff} = \sum_{i=1}^5 C_i \, {Q}_i \, ,
\end{equation}
where the integrated out high-momentum physics is collected into the Wilson coefficients, $C_i$, which 
therefore depend on the underlying theory (SM or BSM). In a commonly used basis the local $\Delta F=2$ 
operators ${Q}_i$ take the form
\begin{equation}
 {Q}_1 =
   \left( \bar{h}\gamma_\mu \,L \, q \right)
   \left( \bar{h}\gamma_\mu \, L \, q \right)\, , \qquad
 {Q}_2 =  \left( \bar{h}\, L \, q \right)
   \left( \bar{h} \, L \, q \right)   \, ,
   \label{eq:Q1}
\end{equation}
\begin{displaymath}
{Q}_3 =  \left( \bar{h}^{\alpha} L \, q^{\beta} \right)
   \left( \bar{h}^{\beta} L \, q^{\alpha} \right) \, , \quad
{Q}_4 =  \left( \bar{h} \, L \, q \right)
   \left( \bar{h}\, R \, q\right) \, , \quad
{Q}_5 =  \left( \bar{h}^{\alpha}  L \, q^{\beta} \right)
   \left( \bar{h}^{\beta} R \, q^{\alpha} \right)\, ,
\end{displaymath}
where $h=b,c,s$ denotes a bottom, charm, or strange quark, $q=d,s,u$ denotes a light (down, strange, or up) quark,
and $R,L = \frac{1}{2} (1 \pm \gamma_5)$. 
The superscripts $\alpha,\beta$ are color indices, which
are shown only when they are contracted across the two bilinears. 
Hadronic matrix elements of these local operators 
are calculable with current lattice-QCD methods with high precision.

The SM prediction for the neutral $B_q$-meson ($q=d,s$) mass difference is given by
\begin{equation}
\Delta m_q = \left( {\rm known} \right) M_{B_q}  |V_{tq}^* V_{tb}|^2 f_{B_q}^2 \hat{B}_{B_q}
\label{eq:Bmix}
\end{equation}
where "(known)" includes short-distance QCD and EW corrections, $M_{B_q}$ is the mass of the
$B_q$-meson, and where $ f_{B_q}^2 \hat{B}_{B_q}$ parameterizes the hadronic matrix element of $Q_1$
in terms of the decay constant $f_{B_q}$ and bag parameter $\hat{B}_{B_q}$
(see, for example, Section 8 of Ref.~\cite{Aoki:2013ldr} for further details). In summary, for neutral $B_q$-meson
mixing lattice QCD needs to provide both the decay constants and the bag parameters, or their combination. 
The ratio of the $B_s$ and $B_d$  mass differences is of particular interest due to the cancellation of statistical
and systematic errors in the corresponding ratio of matrix elements:
\begin{equation}
\frac{\Delta m_s}{\Delta m_d} \, \frac{M_{B_d}}{M_{B_s}} = \left| \frac{V_{ts}}{V_{td}} \right|^2 \, \xi^2 \qquad {\rm with} 
\quad \xi^2 \equiv \frac{f^2_{B_s} \hat{B}_{B_s}}{f^2_{B_d} \hat{B}_{B_d}} \,.
\end{equation}
A similar, but slightly more complicated expression in comparison to Eq.~(\ref{eq:Bmix}) holds for 
the SM prediction of $\epsilon_K$ (see, for example, Section 6 of Ref.~\cite{Aoki:2013ldr}), but only 
the bag parameter $\hat{B}_K$ is needed in this case. For neutral $D$-meson mixing long-distance 
effects in the form of matrix elements of insertions of two effective weak $\Delta C=1$ operators 
are an important, if not dominant, contribution to the SM prediction.  
Similar long-distance effects (albeit with two $\Delta S=1$ insertions) also affect the kaon 
observables $\Delta m_K$ and $\epsilon_K$, where, however, they contribute only a small correction
to the latter observable. 
In the kaon case, first results implementing a strategy for the calculation of such effects have already been 
reported, 
but for $D$ mesons the matrix 
elements of such long-distance operators cannot be calculated with current lattice-QCD
methods. Hence the current goal of lattice-QCD calculations of $D$-meson mixing parameters is not to obtain
information on the CKM parameters, but to help constrain and discriminate between BSM 
theories \cite{Golowich:2007ka}.  

\begin{figure}[t]
    \centering
    \includegraphics[width=0.497\textwidth]{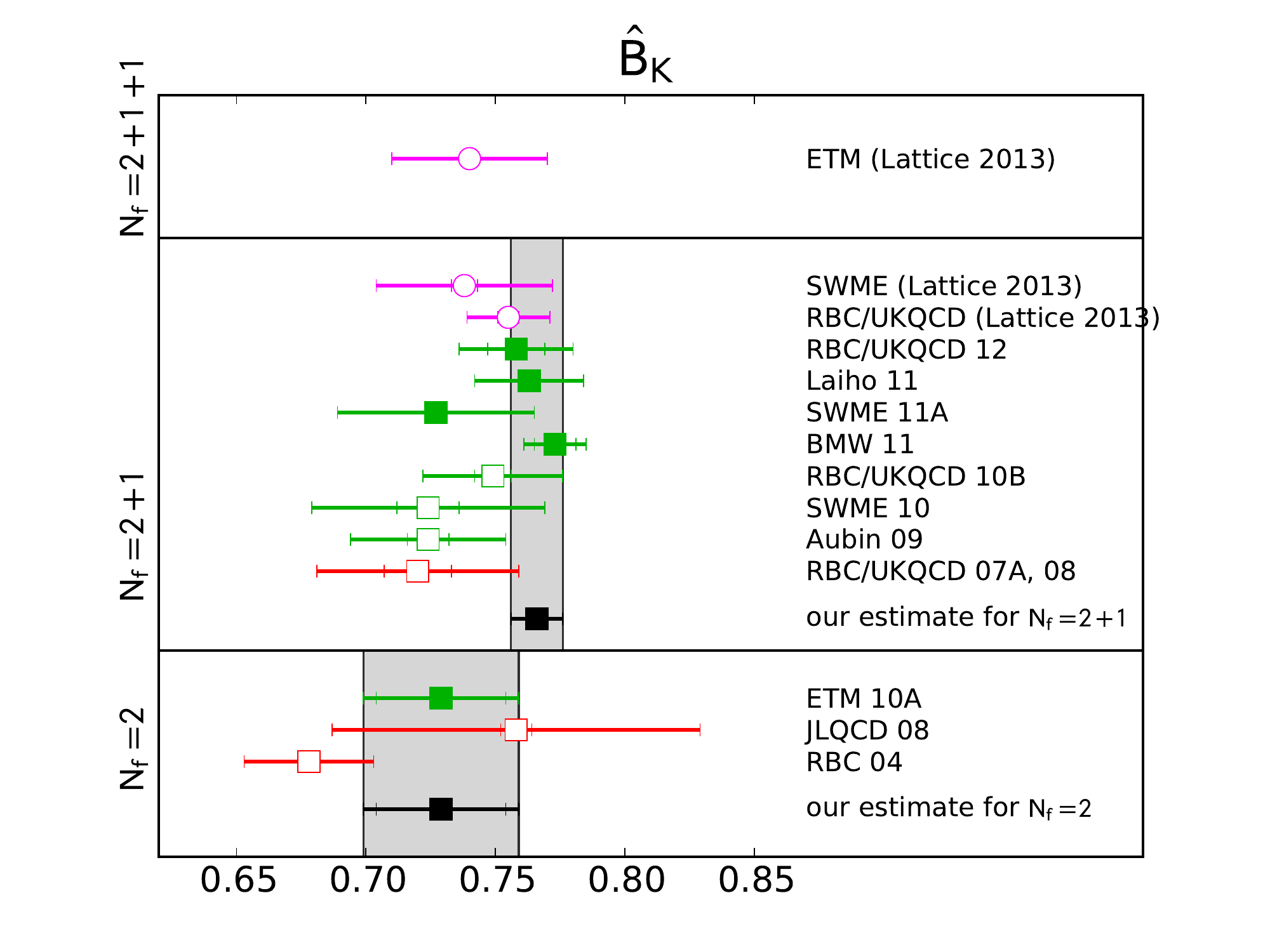} 
     \includegraphics[width=0.497\textwidth]{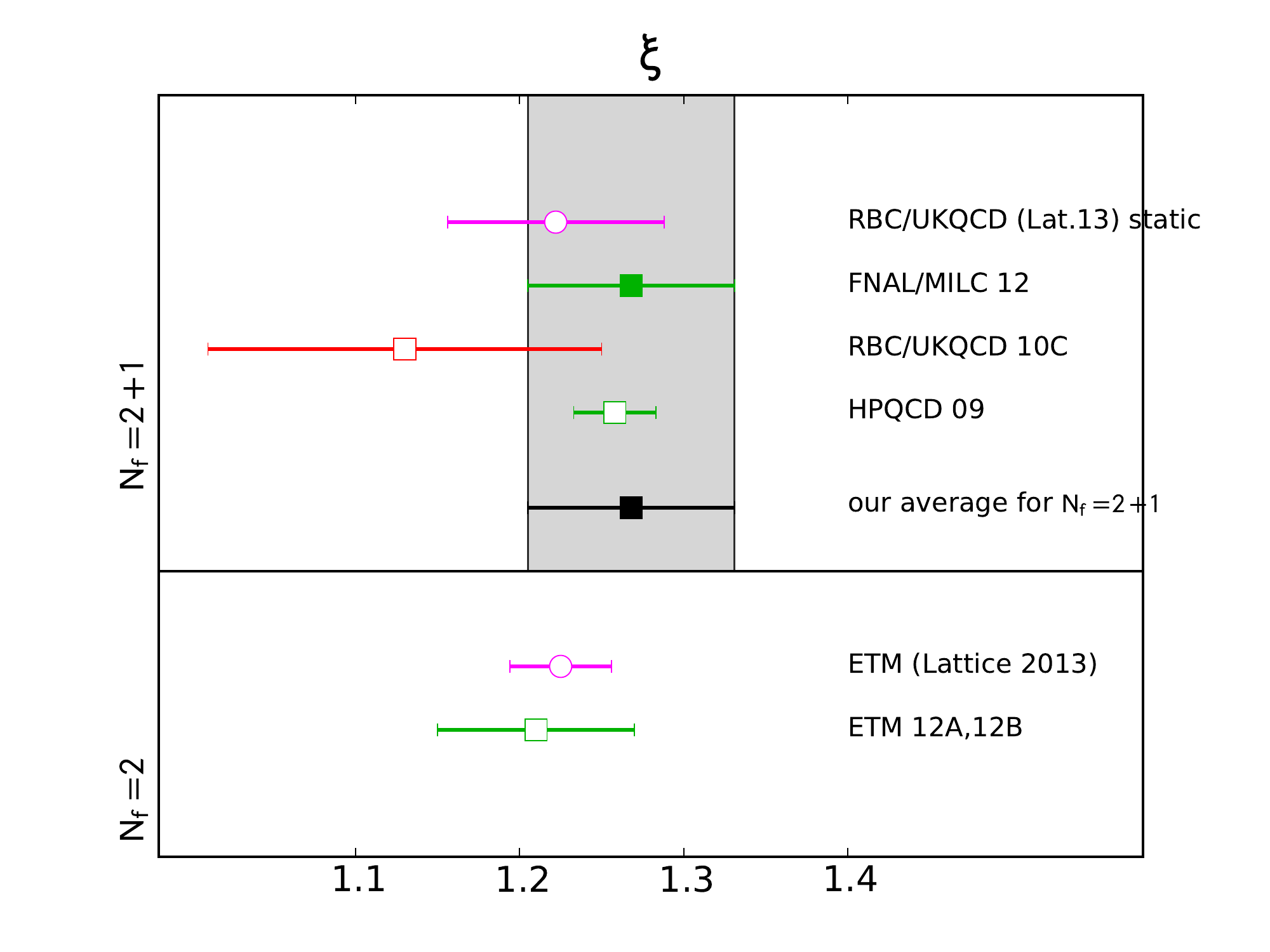}
    \caption{Overview of lattice-QCD results for $B_K$ (left) and $\xi$ (right) adapted from 
    Ref.~\cite{Aoki:2013ldr}. See the caption of Figure~2 for an explanation of the color coding.}
    \label{fig:KBmix}
\end{figure}

Most of the lattice calculations from previous years discussed in the FLAG-2 review have 
focused on hadronic matrix elements of $Q_1$, needed for SM predictions. While it is now common 
practice to calculate the matrix elements of all five operators, recent results for the matrix elements of $Q_{2-5}$ 
are not yet included in the FLAG-2 review. The two panels of Figure~\ref{fig:KBmix}  provide an overview of the 
status of lattice-QCD calculations for $\hat{B}_K$ and for $\xi$. 

The left panel of Figure~\ref{fig:KBmix} shows that a large effort has gone into lattice-QCD calculations of 
$\hat{B}_K$. Four results \cite{Arthur:2012opa,Laiho:2011np,Bae:2011ff,Durr:2011ap} enter the 
FLAG-2 average for $N_f=2+1$, which is quoted with an uncertainty of $1.3$\%,
while one result \cite{Constantinou:2010qv} enters the average for $N_f=2$. Three new results were reported
at this conference, two for $N_f=2+1$ \cite{Bae:2013lja,Frison:2013fga} and one for 
$N_f=2+1+1$ \cite{Carrasco:2013jaa}. 
The first result from physical-mass ensembles was obtained by
the BMW collaboration \cite{Durr:2011ap}. It is the most precise result for $\hat{B}_K$ to date with a 
quoted uncertainty 
of  $1.5$\% and therefore dominates the $N_f=2+1$ FLAG-2 average.  A calculation of $\hat{B}_K$ on 
RBC/UKQCD's new physical mass ensembles was presented at this conference \cite{Frison:2013fga}
with a $2.1$\% uncertainty. The effort of the lattice community to calculate $\hat{B}_K$ has 
already paid off, since its contribution to the $\epsilon_K$ error budget
ranks third, behind the parametric $V_{cb}$ uncertainty and also behind the uncertainty
on the short-distance Wilson coefficient ($\eta_{cc}$) calculated at NNLO in perturbative QCD. In time, we can
expect even smaller errors on $\hat{B}_K$. As already mentioned above, there is also considerable recent
activity for calculations of matrix elements of the other four operators, with three groups reporting 
new results for the additional bag parameters \cite{Carrasco:2013jaa,Bae:2013mja,Lytle:2013oqa}. 

A final comment concerns the relevance of the calculation in Ref.~\cite{Carrasco:2013jaa} of $\hat{B}_K$ 
using ETM's $N_f=2+1+1$ ensembles for the SM prediction of $\epsilon_K$. 
The SM prediction of $\epsilon_K$ in terms of a hadronic matrix element of the local $\Delta S=2$ operator
$Q_1$ is obtained after integrating out the high-momentum degrees-of-freedom, including the charm quarks. 
Indeed, the relatively low matching scale produced by this procedure plays an important role in the relatively
large uncertainty for the corresponding Wilson coefficient ($\eta_{cc}$). With charm quarks present in the 
$N_f=2+1+1$ sea a complete calculation requires the inclusion of the matrix element of the corresponding 
long-distance operator with two insertions of $\Delta S=1$ operators into the calculation. However, as I already 
mentioned, no significant differences between lattice results obtained with different values of $N_f$ have been 
observed at the current level of precision, and $\hat{B}_K$  is no exception. While the inclusion of charm quarks
in the sea is unlikely to have a noticeable effect on $N_f=2+1+1$ results for $\hat{B}_K$ in practice, this 
issue is the reason why these results  are not included in the FLAG-2 review at present.
Finally, I note that for pretty much
every other quantity, calculations that include dynamical charm quarks are ``better'' in the sense that such 
calculations include an effect that is neglected (or included at most perturbatively) in calculations based on
$N_f=2+1$ simulations, however small this effect actually is. 

Since $\hat{B}_K$ is now known with near $1$\% precision, some lattice groups have turned their attention
to the more difficult problems of  calculating nonleptonic decay amplitudes as well as long-distance 
corrections to kaon mixing quantities. Remarkable progress towards understanding the famous 
$\Delta I = 1/2$ rule for nonleptonic kaon decay is evident from the recent results reported at this conference. 
The calculations are based on the Lellouch-L\"uscher method \cite{Lellouch:2000pv}, which requires large physical
volumes and physical pion masses.  The RBC/UKQCD collaboration \cite{Janowski:2013hva}
presented a result  for the 
$\Delta I=3/2$ channel on physical-mass ensembles with Domain-Wall fermions at three different 
lattice spacings. They provide a complete error budget and quote a preliminary total error of an 
impressive $11$\%. Their calculation of the $\Delta I =1/2$ channel is still ongoing. 
In addition, Ref.~\cite{Ishizuka:2013kxa} presents a first attempt of a calculation of the 
$\Delta I=3/2$ and $\Delta I =1/2$
amplitudes using nonperturbatively improved Wilson fermions. Another focus of ``beyond-easy" 
calculations in the kaon system are the long-distance corrections to $\Delta m_K$ and
$\epsilon_K$, for which first results were reported by the RBC/UKQCD collaboration at this 
conference \cite{Christ:2014qaa}.  
  
Turning now to the $B$-meson system, the right panel of Figure~\ref{fig:KBmix} shows that for 
neutral $B$-meson mixing much fewer 
calculations exist. The status for other quantities, such as $B_d$ and $B_s$ bag parameters
and mixing matrix elements is similar to what is shown in Figure~\ref{fig:KBmix} for $\xi$. 
Prior to this conference three results for $N_f=2+1$  had been reported, one
using lattice NRQCD \cite{Gamiz:2009ku}, another based on the static 
limit \cite{Albertus:2010nm}, and the third obtained with Fermilab $b$ 
quarks \cite{Bazavov:2012zs,Bouchard:2011xj}, 
with only Ref.~\cite{Bazavov:2012zs} entering the FLAG-2 average. For $N_f=2$ the ETM collaboration
 \cite{Carrasco:2012de}  
reported at Lattice 2012 on the results of their first calculation of $B$-meson mixing parameters 
using the ratio method. Concerning the new results reported at this conference, 
RBC/UKQCD \cite{Ishikawa:2013faa}  presents an update of  Ref.~\cite{Albertus:2010nm} with the 
inclusion of
ensembles at an additional lattice spacing and also at lighter sea-quark masses. Similarly, 
Ref.~\cite{Carrasco:2013zta} presents the ETM collaboration's final results for $B$-meson mixing
quantities on their $N_f=2$ ensembles. As for kaon and $D$-meson mixing, two 
recent $B$-meson mixing analyses include calculations of hadronic matrix elements of $Q_{2-5}$
\cite{Carrasco:2013zta,Bouchard:2011xj}. 
In summary, while the current status of $B_{(s)}$-mixing calculations is not as good as for decay constants,
the level of recent activity suggests that we can expect significant improvements in the near future.
In particular, given how much room there is for improvement, 
it may be that in the future the best SM prediction for the $B_s \to \mu^+ \mu^-$ decay rate comes
from the indirect method proposed in Ref.~\cite{Buras:2003td}, where the decay rate is normalized
by $\Delta m_s$, and where the lattice input is the $B_s$ bag parameter.  
 
There are several recent efforts for lattice-QCD calculations of the $D$-meson mixing matrix elements
for all five local operators. The ETM collaboration has reported on results obtained with twisted-mass 
Wilson fermions from both their $N_f=2$ \cite{Carrasco:2013zta,Carrasco:2012de} and their $N_f=2+1+1$ 
 \cite{Carrasco:2013jaa} ensembles. An ongoing effort to calculate the complete set of $D$- and $B$-meson
 mixing matrix elements on the full Asqtad set of ensembles with Fermilab bottom and charm quarks is 
 reported by FNAL/MILC in Ref.~\cite{Chang:2013gla}. Studies of the important long-distance correction to 
 SM $D$-meson mixing or of nonleptonic $D$-meson decay require new lattice methods. The first steps 
 towards developing such methods have recently been taken \cite{Hansen:2013dla}. 

\section{Conclusions and Outlook}
\label{sec:con}

The investment made by a growing number of lattice collaborations to produce
ensembles with light quarks simulated at their physical masses and with large physical volumes
is in part responsible for the remarkable progress that has been made in quark flavor physics 
calculations in recent years. The impact of the availability of physical-mass ensembles is most 
apparent in the kaon system, where results for $\hat{B}_K$, $f_K$, and $f_{\pi}$ are reported 
with uncertainties close to or below $1$\% and where the $SU(3)$ breaking quantities 
$f_K/f_{\pi}$ and $f^{K \to \pi}_+ (0)$ are now known to better than $0.5$\%. Remarkably, 
the contribution of $\hat{B}_K$ to the $\epsilon_K$ error budget 
is in third place. Quite appropriately then, a significant effort is now dedicated towards 
calculations of nonleptonic kaon decay amplitudes as well as long distance corrections 
to $\epsilon_K$ and $\Delta m_K$, where physical-mass pions and large physical
volumes are an essential ingredient. First results with complete error budgets at the $10$\%
level have now appeared for the easier case, the $\Delta I =3/2$ amplitude. More
results will no doubt follow soon. 
The contributions from $f_K/f_{\pi}$ and $f^{K \to \pi}_+ (0)$ to the respective
$|V_{us}|$ error budgets will very soon be smaller than those from other sources with the 
simulations and computations that are under way. 
The leading strong (and EM induced) isospin breaking effects are already included in these
calculations. With further improvements the inclusion of sub leading 
corrections via simulations with non-degenerate light sea quarks (using, for example, reweighting 
techniques) and also via adding QED to the simulations are all part of the overall program. 
However, the calculation of the long-distance radiative corrections (which are currently estimated 
using $\chi$PT) with lattice methods requires new methods.  

In the $D$- and $B$-meson systems, results from physical-mass ensembles
have so far been reported only for decay constants. In the next few years physical mass 
ensembles will certainly find their way into calculations of 
$D$- and $B$-meson semileptonic decay form factors and $B$-meson mixing 
quantities. The recently reached sub percent accuracy in $D$-meson decay constant results 
demonstrates the potential effect that physical-mass ensembles can have in reducing the errors
in the charm sector. But even prior to these results, the now widespread use of improved light-quark
actions and the availability of ensembles at small lattice spacings made possible lattice-QCD
results for $D$-meson decay constants and form factors with uncertainties in the range of 
$1-5$\%. 
$B$-meson decay constants have received a lot of attention with a similar (to the $D$-meson case) 
number of independent lattice-QCD results available, all using different heavy-quark methods. 
The present situation for $B$-meson form factors and mixing quantities is much more sparse. However, as 
discussed in Section~\ref{sec:B}, there are many different processes and quantities to consider and for each
process several new results were presented at this conference. 
Given the level of activity, we should see a significant improvement in the reported 
uncertainties for many of these quantities in the next few years. This is good news since the $B$-meson 
decay constants, semileptonic form factors, and mixing parameters are important inputs to CKM fits. 
Already, lattice calculations of several $D$- and $B$-meson quantities have reached the 
level of precision where isospin breaking and EM effects are becoming relevant. 
Strong (and EM induced) isospin breaking effects are straightforward to include, the same as for kaon decays. 
However, quantitative phenomenological predictions of structure dependent EM effects in 
(semi)leptonic $D$- and $B$-meson decays similar to the ones for kaon decay \cite{Cirig:2011xxx} 
(including an error budget), don't currently exist. Such calculations are now needed.  

\begin{figure}[t]
    \centering
    \includegraphics[width=0.6\textwidth]{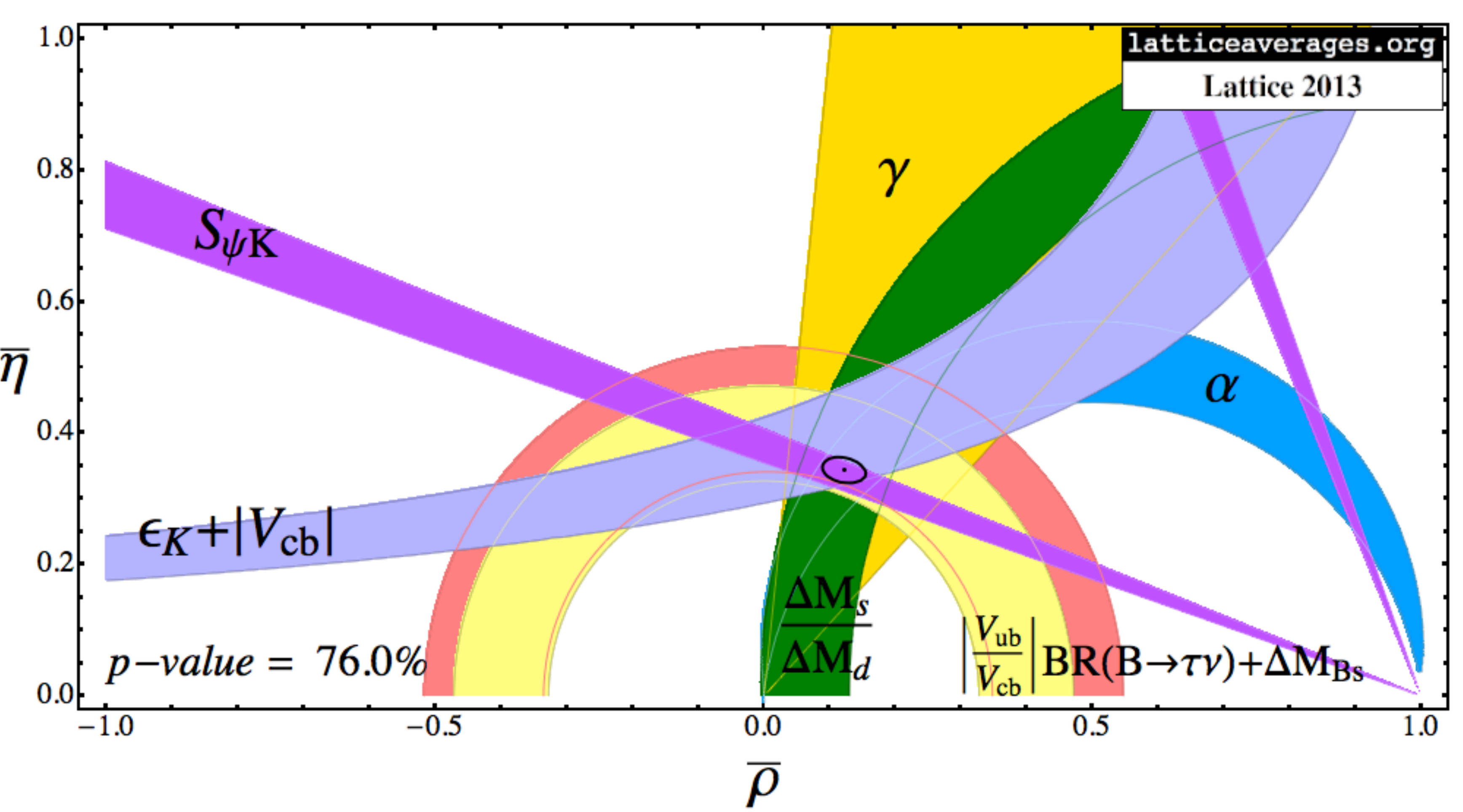} 
     \caption{Example of a UT fit obtained in Refs.~\cite{Laiho:2009eu,Laiho:2012ss} using the FLAG-2 averages from 
     Ref.~\cite{Aoki:2013ldr}. See Refs.~\cite{Laiho:2009eu,Laiho:2012ss} for more details.}
    \label{fig:utfit}
\end{figure}

With the FLAG-2 averages in hand, one can now consider how the lattice results affect the SM CKM picture. 
This is done in a global analysis that takes the relevant averages of experimental measurements together with
the FLAG-2 averages for the relevant hadronic parameters as inputs and performs a fit to the SM CKM 
parameters, in particular, to obtain constraints on $\rho$ and $\eta$. Such fits are usually referred to as unitarity 
triangle fits. There are currently three groups 
which perform such analyses: LLV \cite{Laiho:2009eu,Laiho:2012ss}, the UTfit collaboration \cite{Bevan:2012waa},
and the CKM fitter group \cite{ckmfit}. There are some differences in which inputs are used between the
three groups, in particular which lattice results are used, and how the errors are treated. Figure~\ref{fig:utfit}
shows an example of a unitarity triangle fit obtained from LLV, which I discuss here
since it uses the FLAG-2 averages of Ref.~\cite{Aoki:2013ldr} for 
$\hat{B}_K$, $f_K$, $\xi$, $f_{B_s} \sqrt{\hat{B_{B_s}}}$,  $f_B$, $|V_{ub}|_{\rm excl}$, and 
$|V_{cb}|_{\rm excl}$. The exclusive $|V_{u(c)b}|$ are averaged with the inclusive ones using the PDG
prescription to inflate the errors when combining inconsistent results.  
See Refs.~\cite{Laiho:2009eu,Laiho:2012ss} for a list of the other inputs.  
Since the unitarity triangle parameters $\rho$ and $\eta$ are overconstrained with all the inputs, one 
can remove constraints and instead obtain predictions for them. The resulting differences provide 
information in addition to the goodness of fit ($p$-value) of how consistent the various inputs are with each other. 
If there are inconsistencies, one can then also use this analysis to anticipate where new physics may most 
likely make an appearance. Unfortunately, however, there currently are no big inconsistencies, 
unlike a few years ago. Nevertheless, the persistent $3 \sigma$ 
differences between inclusive and exclusive $|V_{ub}|$ and $|V_{cb}|$ determinations are still 
cause for concern. 

\section*{Acknowledgements} 
Many people have helped me to prepare and write this review. I am grateful to all my colleagues in the FLAG collaboration 
for their work in putting together the review of lattice results relevant to quark flavor physics 
and for all their contributions to this review. In particular I thank my collaborators in the FLAG Heavy Quark working 
groups, Yasumichi Aoki, Michele Della Morte, Enrico Lunghi, Carlos Pena, Junko Shigemitsu, and 
Ruth Van de Water for a thoroughly enjoyable collaboration and for many discussions and insights, which helped to 
shape this review. Furthermore,
I thank  Mariam Atoui, Jon Bailey,  Claude Bernard, Fabio Bernardoni, Benoit Blossier, Chris Bouchard, Nuria Carrasco, 
Yi-Bo Chang, Ting Wai Chiu, Christine Davies, Carleton DeTar, 
Petros Dimopoulos, Rachel Dowdall, Daping Du, Elvira G\'amiz, Jochen Heitger, Tomomi Ishikawa, Jonna Koponen, 
Andreas Kronfeld, Jack Laiho, Weongjong Lee, Vittorio Lubicz, Bob Mawhinney, 
Paul Mackenzie, Eleonora Picca, Lorenzo Riggio, Francesco Sanfilippo, Jim Simone, Amarjit Soni, Doug Toussaint, 
Oliver Witzel, and Ran Zhou for providing me with information and/or for discussions relevant to this review. 
Finally, I thank the Fermilab Theory Group for hospitality while this review was 
written. This work was supported in part by a URA Fermilab Visiting Scholarship and by the Department of Energy under
grant no.~DE-FG02-13ER42001.


\end{document}